\documentclass[%
 reprint,
 amsmath,amssymb,
 aps,
 nolongbibliography,
prl,
]{revtex4-2}

\usepackage{graphicx}% Include figure files
\usepackage{relsize}
\usepackage{placeins}

\usepackage{bm}% bold math
\usepackage{amsmath}
\usepackage{xcolor}
\usepackage{ulem}

\usepackage{float}
\usepackage{graphicx}
\usepackage{hyperref}
\usepackage{amsmath}
\usepackage{ amssymb }
\usepackage{fontawesome}
\usepackage{cancel}
\usepackage{soul}
\usepackage{ulem}

\begin{document}

\newcommand{\wjr}[1]{\textcolor{red}{#1}}
\newcommand{\sea}[1]{\textcolor{black}{#1}}
\newcommand{\sear}[1]{\textcolor{black}{#1}}
\newcommand{\searr}[1]{\textcolor{black}{#1}}

\setlength{\abovedisplayskip}{2pt}
\setlength{\belowdisplayskip}{2pt}

\preprint{APS/123-QED}

\title{Boundary-driven delayed-feedback control of spatiotemporal dynamics in excitable media}

\author{Sebasti\'an Echeverr\'ia-Alar}
\author{Wouter-Jan Rappel}%

\affiliation{Department of Physics, University of California, San Diego, California 92093, USA}

%\date{\today}
 
\begin{abstract}
Scroll-wave instabilities in excitable domains are central to life‐threatening arrhythmias, yet practical methods to stabilize these dynamics remain limited. Here, we investigate the effects of boundary layer heterogeneities in the spatiotemporal dynamics of a quasi-2D
semidiscrete excitable model. We reveal that a novel boundary-driven mechanism
suppresses meandering and chaotic spiral dynamics. We show how the strength of the heterogeneities mediates the emergence of this regulation through a pinning-unpinning-like transition. We derive a reduced 2D
model and find that a decrease in bulk excitability and a boundary-driven delayed-feedback underlie the stabilization. Our results may point to alternative methods to control arrhythmias.
\end{abstract}

\maketitle

Spatially extended nonlinear systems can exhibit complex spatiotemporal dynamics, often after undergoing an  instability   \cite{cross1993pattern,cross2009pattern}.  Examples include signaling protein waves on cell membranes \cite{tan2020topological}, oscillations in chemical reactions \cite{zaikin1970concentration, ouyang2000transition}, Rayleigh-B\'enard convection patterns \cite{maurer1979rayleigh,egolf2000mechanisms,cross1995domain}, turbulence in active matter \cite{martinez2019selection}, extreme events in lasers \cite{pammi2023extreme}, and rogue waves in dissipative optical systems \cite{tlidi2022rogue}. One  question that has attracted considerable attention is how this complexity can be controlled and how 
these instabilities can be suppressed \cite{walkama2020disorder,ivars2022optical,beppu2024magnetically}.
This is a particularly relevant question  
in the context of cardiac dynamics. Cardiac tissue is a prototypical example of an excitable medium, where cardiomyocytes are interconnected via gap junctions \cite{rappel2022physics}. 
Under normal conditions, a planar electrical activation front sweeps through the tissue, but under abnormal circumstances this planar wave can become unstable and the break-up of the front can result in the formation of spiral waves. This disorganized behavior leads to serious 
 heart rhythm disorders, including tachycardia and fibrillation \cite{panfilov2002spiral,karma2013physics,rappel2022physics}.
 Spiral wave self-organization, however, is not limited 
 to cardiac tissue and is a common feature of excitable media, observed in the brain cortex \cite{huang2010spiral,xu2023interacting}, multicellular aggregates \cite{sawai2005autoregulatory}, bacterial populations \cite{liu2024emergence}, and cytoskeletal filaments \cite{bourdieu1995spiral}.

In cardiac tissue slices, the  simplest form of a spiral wave rotates around a circular core. 
Changing the excitability properties of the tissue can destabilize this rigidly rotating spiral wave, resulting in a meandering spiral \cite{barkley1990spiral,barkley1992linear,li1996transition} or in wave break-up, leading to spiral defect chaos \cite{fenton2002multiple,vidmar2019extinction}.
Cardiac tissue, has finite thickness and 
3D effects may play an important role in the formation of instabilities.
The 3D equivalent of a spiral wave is a scroll wave: a stack of 2D spiral waves whose tips form a filament  \cite{winfree1973scroll,pismen1999vortices}. In sufficiently thick geometries, filaments add degrees of freedom  and their dynamics is  richer in both homogeneous \cite{keener1988dynamics,pertsov1990three,keener1992dynamics,pismen1999vortices,henry2002scroll}, and heterogeneous tissues \cite{alonso2011effects,ten2005wave,ten2007influence}.
Conversely, in thin, scroll waves exhibit dynamics similar to the spiral waves in 2D \cite{alonso2013negative}.

Successful control of scroll wave instabilities in cardiac tissue has obvious
clinical significance and can prevent or even revert tachycardia and fibrillation to normal sinus rhythm  \cite{rappel1999spatiotemporal,luther2011low}. Yet a standard, minimally invasive technique remains debated. Here we demonstrate that introducing boundary layer heterogeneities in a thin computational tissue slab can increase scroll wave stability. 
Specifically, we show that
decreasing the coupling strength between cardiac cells or their excitability properties near the slab boundaries can stabilize scroll waves in the bulk against meandering and break-up instabilities, and even terminate wave activity. Moreover, we theoretically establish a boundary-driven delayed-feedback mechanism underlying the stabilization.

\begin{figure}[h!]
\makeatletter
\typeout{FIGSIZE1: \the\Gin@nat@width\space x \Gin@nat@height}
\makeatother
\includegraphics [width=6.6 cm]{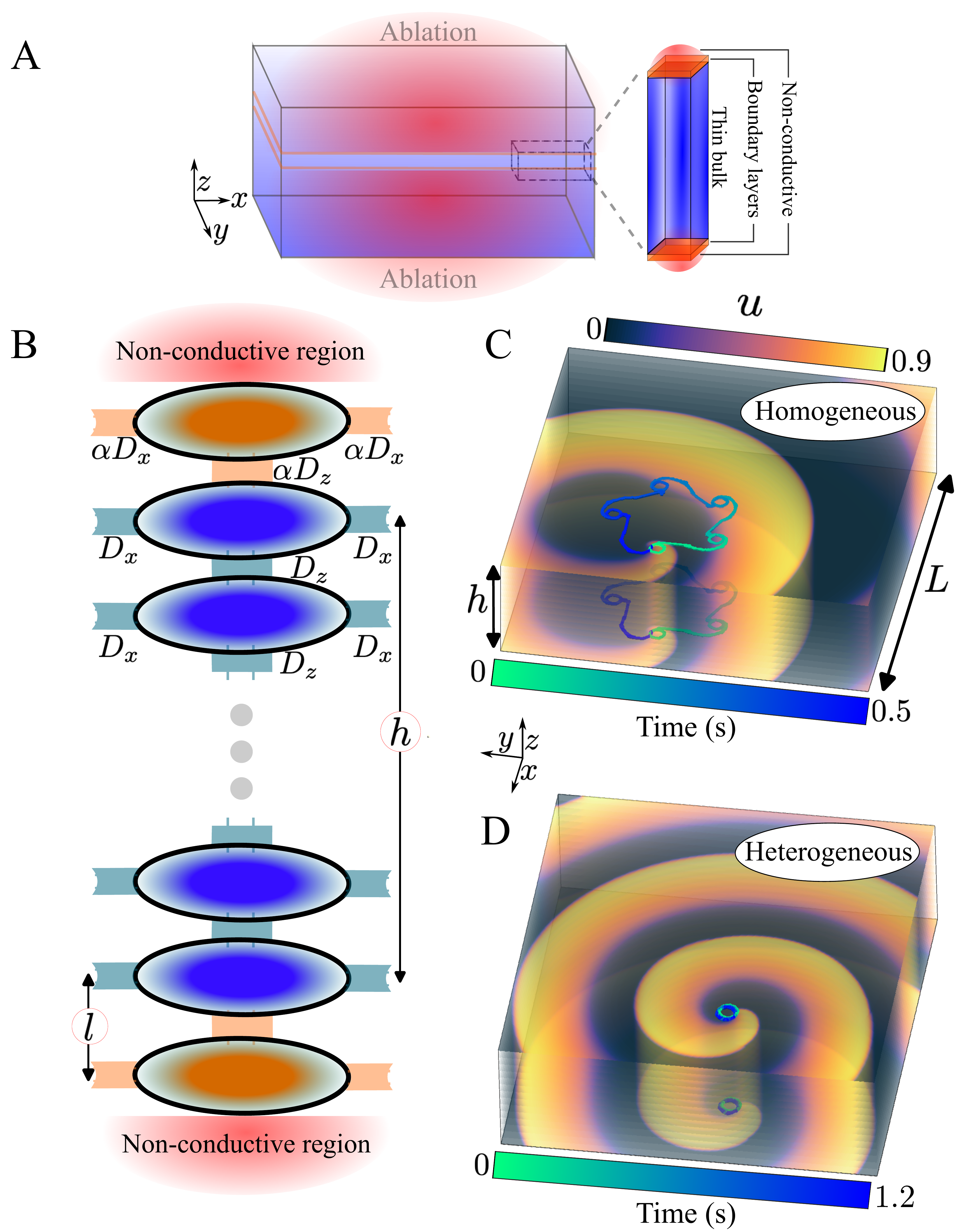} 
\caption{Boundary layer effects. (A) Cartoon of a non-transmural ablation. (B) Schematic representation of the discrete coupling between cardiomyocytes along the transmural direction. (C) Star-like tip trajectories and the variable $u$ in the top and bottom slice for the homogeneous case at $\tau_{d}=0.382$ ms, $\mathcal{H}=14d_{z}$. (D) As in B with boundary 
layer heterogeneities of strength $\alpha=0.004$ and size $d_{z}$.}
\label{Fig01_observation}
\end{figure}

The choice behind boundary layer heterogeneities is motivated by the fact that arrhythmias are often 
treated by ablation during which tissue is locally destroyed by applying heat, cold, or pulsed electric fields. However, ablation often leads to incomplete lesions, and only renders tissue non-conductive or non-excitable up to a certain depth \cite{bugge2005comparison,jiang2019feasibility,rogers2020continuous,nakagawa2024effects}, thus creating a significant transmural heterogeneity in which a region 
 with impaired conduction (boundary layer) is sandwiched between conductive (bulk) and non-conductive tissue. Ablation lesions are typically asymmetric in the transmural direction, but can also be symmetric (Fig.~\ref{Fig01_observation}A). We concentrate on the latter, as the former can be inferred by symmetry arguments (see Supplemental Material (SM) \cite{SupMat}). Furthermore, following previous numerical
 	studies \cite{zahid2016feasibility,boyle2019computationally,sakata2024assessing}, we only consider   short-term post-ablation effects  and neglect possible long-term
 	remodeling and recovery of   tissue  \cite{AcuteNote}.
 
We consider an electrically insulated tissue slab of thickness $\mathcal{H}\ll L$, where $L= 5.08$ cm is the width of the slab. The slab is divided into a boundary layer of size $l$ and a bulk region of size $h/2$ along the transmural direction $z$ (Fig.~\ref{Fig01_observation}B). We model the boundary layer as a region where the coupling strength between cardiac cells is smaller than in the bulk by a factor $\alpha$. Modulating the coupling strength, rather than excitability, provides a more model-independent approach. 
Nevertheless, as we detail in the SM \cite{SupMat}, altering excitability properties leads to qualitatively similar results. We discretize the $z$-direction in slices with a thickness corresponding to the size of a cardiomyocyte: $d_{z} = 25$ $\mu$m \cite{hsieh2006endothelial}.

To model the excitable dynamics of cardiac cells, we employ the following semi-discrete equation;
\begin{equation}
\partial_{t}u = \mathcal{I}_{k}D_{o}\nabla_{\perp}^2 u + \dfrac{D_{z}}{d_{z}^{2}}\mathcal{C}_{z} - I_{ion} ,\ \ k=\{0,..,\mathcal{H}/d_z\}, 
\label{FKmodel}
\end{equation}
 where $u$ represents the transmembrane voltage in each cell. The first term describes the coupling in the $x\!-\!y$ plane, modeled by isotropic diffusion with a coefficient $D_{o}$, and $\mathcal{I}_{k}$ encodes the heterogeneous coupling strength distribution in the different $k$-slices. The second term describes the discrete coupling along $z$ with a coupling strength $D_{z}/d_{z}^{2}$. Both $\mathcal{I}_{k}$ and $\mathcal{C}_{z}$ depend on $\alpha$ at the boundary layers  (SM \cite{SupMat}, Section I). The last term represents the 
ion currents and is taken from the Fenton-Karma model \cite{fenton1998vortex} (parameters in SM \cite{SupMat}, Section I).

We first investigate how boundary layer heterogeneities can stabilize the meandering instability  in the geometry shown in Fig.~\ref{Fig01_observation}. For this, we choose parameter values for which in the homogeneous case ($\alpha = 1$) the model Eq.~(\ref{FKmodel}) exhibits a scroll wave that rigidly rotates with a frequency $\omega_{o}$ exhibiting a straight filament \cite{tipNote}. Decreasing $\tau_{d}$, a parameter that controls the excitability, the scroll wave starts to  rotate faster and eventually undergoes a meandering instability at a critical value $\tau_{d}^{c}=0.384$ ms \cite{bifurcationNote}. As a result, the filament repeats a star-shaped trajectory every $0.5$ s as evident from the filament trajectories at the top and bottom slices of the slab (Fig.~\ref{Fig01_observation}C).  

Introducing a boundary layer heterogeneity of size $l=d_z$ and strength $\alpha=0.004$ reduces $\tau_{d}^{c}$, increasing the stability range of rigidly rotating scroll waves (Fig.~\ref{Fig01_observation}D and Video S1). We note that  the heterogeneity  need not encompass the entire boundary slice to suppress  meandering and can also stabilize scroll waves in the presence of rotational transmural anisotropy and in thin curved geometries
  (Fig.~S1 and Section II-IV in SM \cite{SupMat}). 

\begin{figure}[h!]
\makeatletter
\typeout{FIGSIZE2: \the\Gin@nat@width\space x \Gin@nat@height}
\makeatother
\includegraphics [width=8.3 cm]{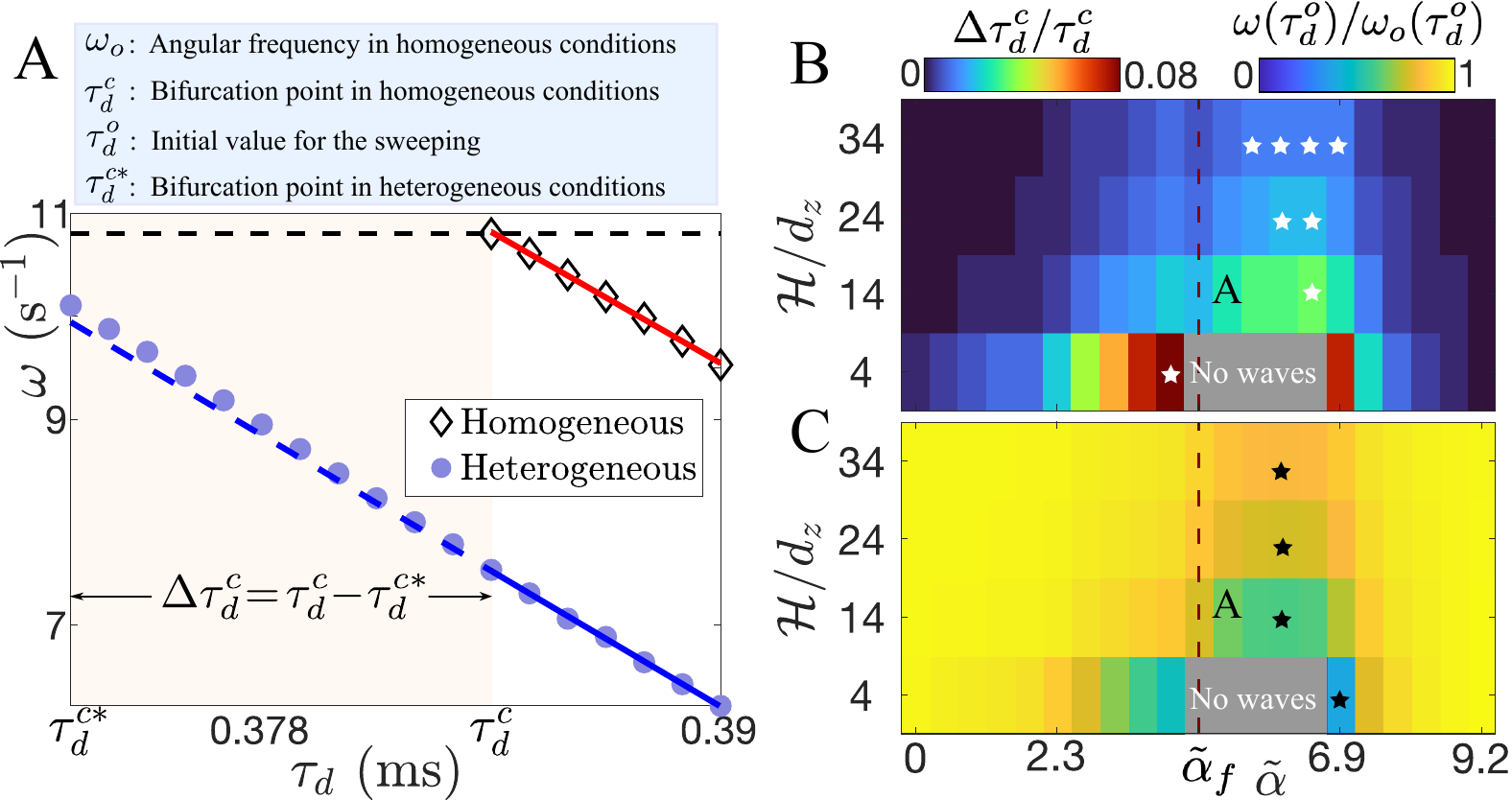}
\caption{Control of meandering instability with $l=d_{z}$. (A) $\omega$ vs $\tau_d$ in homogeneous and heterogeneous conditions. The horizontal black dashed line emphasizes $\omega_{o}(\tau_{d}^{c})$. The solid lines indicate linear fits. (B-C) Phase diagrams in $\tilde{\alpha}-\mathcal{H}/d_{z}$ space showing $\Delta\tau_{d}^{c}/\tau_{d}^{c}$ and $\omega/\omega_{o}$ at $\tau_{d}^{o}$, respectively. Here $\tilde{\alpha}=log(\alpha)+c$ with $c=10.82$ \cite{logNote}. The white (black) stars indicate $\max(\Delta\tau_{d}^{c}/\tau_{d}^{c})$ ($\min(\omega/\omega_{o})$) for each $\mathcal{H}$.}
\label{Fig02}
\end{figure}

To systematically explore the stabilization of scroll waves in the presence of boundary layer heterogeneities of size $d_{z}$, we start with a rigidly rotating scroll wave at $\tau_{d}^{o}=0.390$ ms. We then decrease this parameter every 15 rotations in steps of $0.001$ ms until we trigger the meandering instability. We first carry out this parameter sweep in the absence of a 
heterogeneity, which shows that the frequency $\omega$ increases 
linearly as $\tau_{d}$ decreases (open symbols and red line,  Fig.~\ref{Fig02}A).
Repeating this procedure in the presence of the heterogeneity reveals a slowing down of the scroll wave and 
a shift of the meandering bifurcation, $\tau_{d}^{c*}=0.373$ ms (closed symbols in Fig.~\ref{Fig02}A). Note, however, that the linear dependence between $\omega$ and $\tau_{d}$ is almost conserved in the heterogeneous case, but with a slight deviation within the range $[\tau_{d}^{c*},\tau_{d}^{c}]$, as indicated by the blue dashed line in Fig.~\ref{Fig02}A.

Next, to elucidate the role of the slab thickness and the strength of the coupling at the boundaries in  scroll wave stabilization, we systematically vary $\mathcal{H}$ and $\alpha$ and 
determine $\tau_{d}^{c*}$ and the angular frequency at $\tau_{d}^{o}$. The results can be summarized in two phase diagrams. 
The first one shows the increase in stability $\Delta\tau_{d}^{c}/\tau_{d}^{c}$, where $\Delta\tau_{d}^{c}=\tau_{d}^{c}-\tau_{d}^{c*}$ (Fig.~\ref{Fig02}B). 
The second one displays  the frequency ratio $\omega/\omega_{o}$ at $\tau_{d}^{o}$ (Fig.~\ref{Fig02}C), which is a signature of the boundary layer effect on the scroll wave motion. For a fixed $\mathcal{H}$, both quantities exhibit a non-monotonic behavior, as expected from the equivalence between $\alpha\!=\!0$ and $\alpha\!=\!1$ in our model.

The phase diagrams illustrate that an enhancement of stability is correlated with a reduction in the rotational frequency of the waves. This enhancement becomes larger for thinner slabs, 
which can be expected since the boundary layer becomes proportionally bigger. Additional computational simulations show that larger boundary layers ($l>d_z$) increase bulk stabilization for almost all $\alpha$ values (Fig.~S3 in SM \cite{SupMat}). This 
shows that to observe a significant $\Delta\tau_{d}^{c}/\tau_{d}^{c}$, and the corresponding decrease in $\omega$, it is not  necessary to reduce the coupling strength by a factor of 100-500. 
Instead, a  reduction that is almost two orders of magnitude smaller can be sufficient if the boundary layer size is large enough. At the same time, larger boundary layers allow a substantial $\Delta\tau_{d}^{c}/\tau_{d}^{c}$ in thicker slabs (Fig.~S3B in SM \cite{SupMat}), demonstrating that the stabilization mechanism is robustly present across slab thicknesses with a suitable chosen $l$ \cite{HHnote}. Note that the ratio $2l/\mathcal{H}$ measures the amount of tissue with reduced conductivity, and its impact on the reduction of $\omega$ is similar to what has been observed in 2D reaction-diffusion models with  randomly distributed weakly conductive inclusions \cite{alonso2009effective}. A further consequence of this ratio is that the degree of stabilization is conserved when considering only one heterogeneous boundary in a slab of thickness $\mathcal{H}/2$ (Fig.~S4 in SM \cite{SupMat}).

A relevant division in the phase diagrams occurs at $\alpha_{f}=0.002$ (dashed line in Figs.~\ref{Fig02}B-C). For $\alpha<\!\!\alpha_{f}$, wave propagation from the bulk fails to excite the boundaries, while for $\alpha\!\!\geq\!\!\alpha_{f}$, it succeeds. In the gray areas in Figs.~\ref{Fig02}B-C, the introduction of the boundary layers terminates wave activity in the bulk and renders the tissue unexcitable. This behavior is also observed when the boundary layers are thicker (Fig.~S3 in SM \cite{SupMat}).

\begin{figure}[h!]
\makeatletter
\typeout{FIGSIZE3: \the\Gin@nat@width\space x \Gin@nat@height}
\makeatother
\includegraphics[width=7.6 cm]{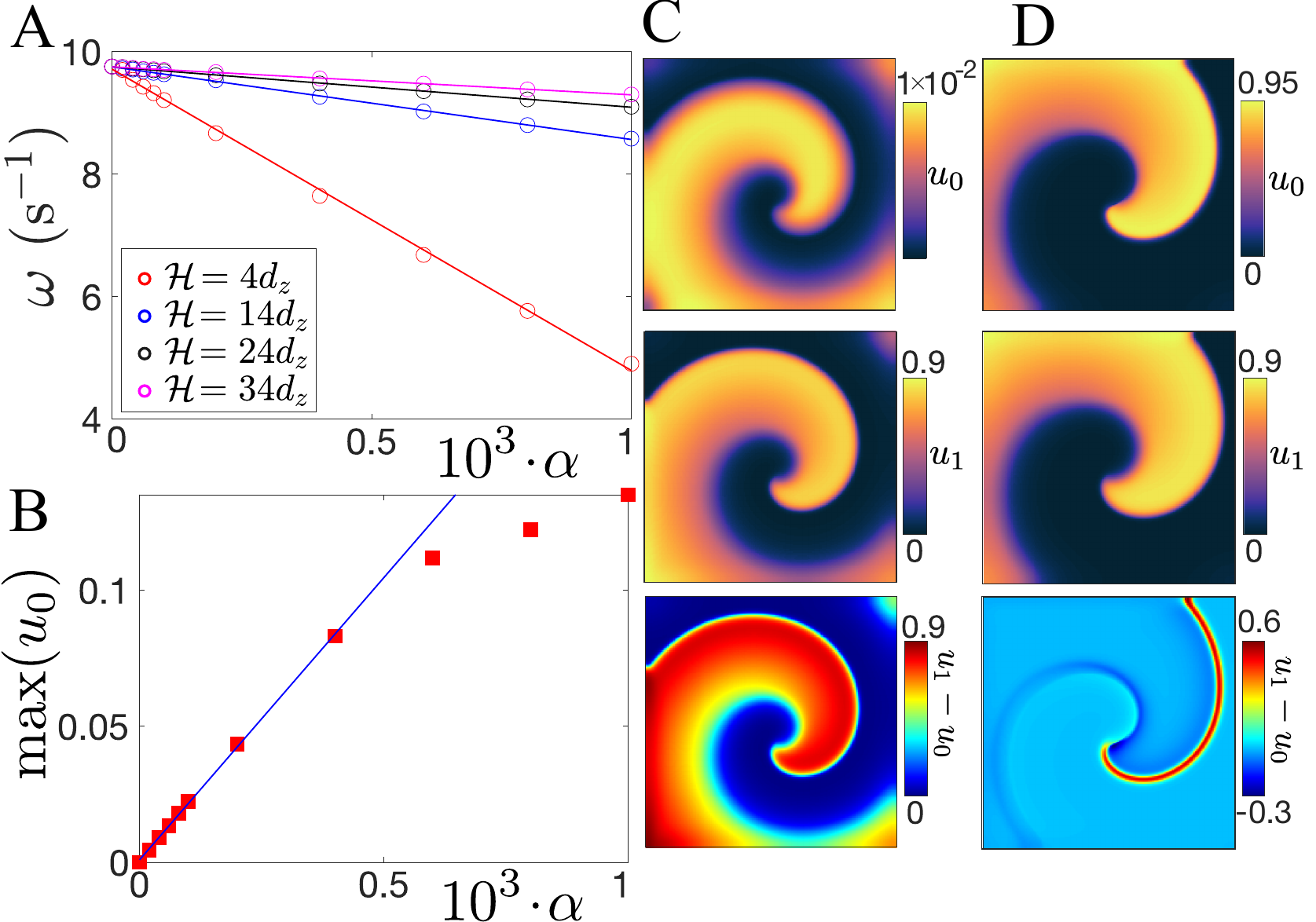}
\caption{Dynamical regimes of scroll waves. (A) $\omega$ vs $\alpha$ for different $\mathcal{H}$ in the leakage regime. (B) $\max(u_{0})$ vs $\alpha$ ($\mathcal{H}=14d_{z}$ and $l=d_{z}$). (C-D) $u_{o}$, $u_{1}$ and $u_{1}-u_{o}$ in the (C) leakage ($\alpha=4\!\times\!10^{-5}$) and (D) feedback ($\alpha=4\!\times\!10^{-3}$) regimes.}
\label{Fig03}
\end{figure}

To gain insight into the suppression of the meandering instability, we first analyze separately the cases $\alpha\!\!<\!\!\alpha_{f}$ and $\alpha \!\!\geq \!\!\alpha_{f}$. In the former case,  $\omega$ varies linearly as a function of 
$\alpha$ for all  values of $\mathcal{H}$ (Fig.~\ref{Fig03}A). Furthermore, in this case,  the maximum of the 
potential in the boundary slice, $u_0$, increases linearly with $\alpha$, until $\alpha = 5\!\times\!10^{-4}$ (Fig.~\ref{Fig03}B), but is significantly lower that the maximum value in the bulk slice next to the boundary, $u_{1}$
($\approx 0.92$). We therefore call this the \textit{leakage} regime: wave propagation in the bulk is not enough to fully excite the boundary slices and the potential in the bulk leaks
into the boundary layers (Fig.~\ref{Fig03}C).  The propagation failure of the action potential into the boundary layers  resembles wave pinning in discrete systems governed by excitable or bistable dynamics \cite{rinzel1990mechanisms,fath1998propagation,mitkov1998tunable,carpio2001wave,clerc2011continuous}, and the growth of $\max(u_{0})$ can be interpreted as a precursor to wave unpinning. 

The linear decrease of $\omega$, as a function of $\alpha$, can be analytically addressed in the case $\alpha\!\!<\!\!\alpha_{f}$. For this, we consider a tissue slab of size $\mathcal{H}=2d_{z}$, i.e., one bulk slice sandwiched by two boundary layers. Since $u_{0}\!\!\ll\!\! u_{1}$, the dynamics can be analyzed by considering only the bulk slice, where $C_{z}=-2\alpha D_{z}/d_{z}^{2}u_{1}+\mathcal{O}(\alpha u_{0})$. 
This negative linear term implies a reduction of excitability in the bulk. Similar to previous works \cite{setayeshgar2001scroll,henry2002scroll}, we perform a weakly nonlinear analysis and find that the $\mathcal{O}(\alpha)$ correction in $\omega$ is $\Delta\omega\propto-2D_{z}\alpha/d_{z}^{2}$ (SM \cite{SupMat}, Section VII). Consistent with our numerical results, this analysis shows that  the decrease of excitability  in the bulk, governed by the boundary layer heterogeneities, reduces $\omega$. Although this analysis is only valid for a non-meandering spiral wave in the homogeneous case, i.e. $\tau_{d}\!>\!\tau_{d}^{c}$, our numerical results support the idea that a reduction in $\omega$ plays a role in the delay of the meandering instability (Figs.~\ref{Fig02}B-C). 
%and the eventual excitation of the boundary layer with increasing $\alpha$

When $\alpha\ge\alpha_f$, an abrupt unpinning-like transition lets bulk waves fully excite the boundary layers (SM \cite{SupMat}, Sec.~VIII), switching to a feedback regime where the boundary regulates the bulk and a bulk scroll drives a boundary spiral, producing a nontrivial voltage difference
	at the bulk-boundary layer interface (Fig.~\ref{Fig03}D).
	 We note that the boundary layers usually cannot excite the bulk on their own. This type of nonreciprocal propagation failure has been studied in 1D and 2D reaction-diffusion models \cite{ikeda1989wave,rudy1995reentry,kulka1995influence,zykov2017fast}, where it was shown that sharp jumps in conductivity are the main cause of this behavior. In our case, however, this is not the whole story as the ratio $2l/\mathcal{H}$ controls the ability of the boundaries to excite the bulk. When $2l/\mathcal{H}\lesssim 0.5$, the boundaries fail to excite the bulk even as $\alpha\rightarrow 1$, and when $2l/\mathcal{H}\gtrsim 0.5$, the boundaries can activate the whole slab in the feedback regime.
% Note that there is a two order of magnitude difference between $u_{0}$ and $u_{1}$ in the leakage regime.

\begin{figure}[h!]
\makeatletter
\typeout{FIGSIZE4: \the\Gin@nat@width\space x \Gin@nat@height}
\makeatother
\includegraphics[width=8.5 cm]{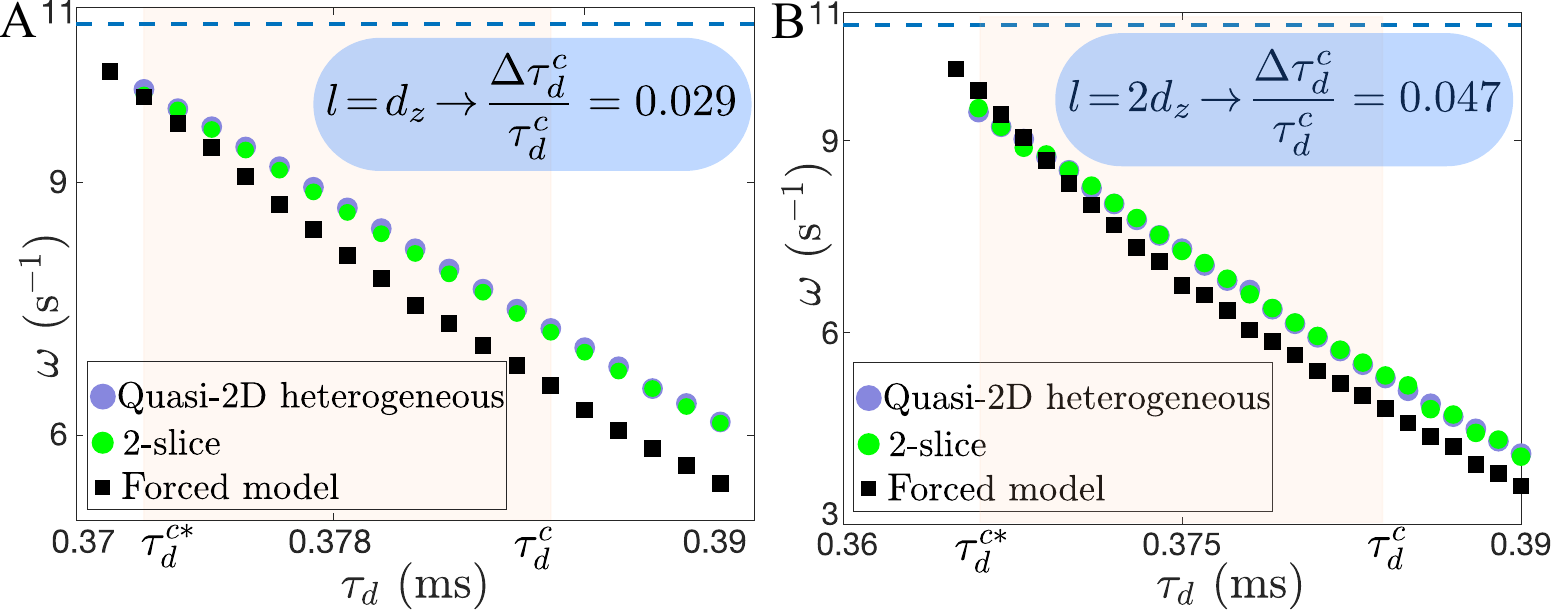}
\caption{$\omega$ vs $\tau_{d}$ in the different models for the case $\mathcal{H}=14d_{z}$ and $\alpha =0.005$. (A) $l\!=\!d_{z}$, (B) $l\!=\!2d_{z}$. The dashed line corresponds to the $\omega_{o}(\tau_{d}^{c})$, and the shaded area to $\Delta\tau_{d}^{c}$.}
\label{Fig04}
\end{figure}

When $l=d_{z}$, the problem can be simplified by deriving a 2-slice model,  consisting of only a boundary layer slice and 
a bulk slice. For this, we use the fact that 
for all values of $\alpha$ and $h$, our simulations show that in the bulk the term $\mathcal{C}_{z}=D_{z}\partial_{zz}u$ is approximately independent of $z$ \cite{continuumNote}.
Then, in the rotational frame of reference,  we use the symmetry of our geometrical set-up and solve the Laplace equation  in one half of the slab with the Dirichlet boundary condition $u=u_{m}$ at $z=0$, where $u_{m}=u_{\mathcal{H}/2dz}$ is the transmembrane voltage at the middle slice, and the Neumann boundary condition $\partial_{z}u = \alpha(u_{0}-u_{1})/d_{z}$ at $z=h/2$. The solution, back in discrete form and for a general size $l$, is $u_{k}=u_{m}+\chi_{k}(u_{b}-u_{m})$, where $\chi_{k}=4\alpha[kd_{z}-l-h/2]^{2}/(4hd_{z}+\alpha h^{2})$ with
$k$ running from $k=l/d_{z}$ to $k=(\mathcal{H}-l)/d_z$, and $u_{b}=u_{l/d_{z}-1}$ is the boundary transmembrane voltage next to the bulk. The $u_{k}$ fields can be introduced in Eq.~(\ref{FKmodel}), reducing the full quasi-2D equations into a far simpler 2-slice model
\begin{equation}
\fontsize{9.4}{12}\selectfont
\begin{split}
\partial_{t}u_{m} &= D_{o}\nabla^{2}_{\perp}u_{m} - I_{ion}^{m} +2\dfrac{D_{z}}{d_{z}^{2}}\chi_{m+1}(u_{b}-u_{m})\\
\partial_{t}u_{b} &= \alpha D_{o}\nabla^{2}_{\perp}u_{b} - I_{ion}^{b} + \alpha\dfrac{D_{z}}{d_{z}^{2}}(1-\chi_{b+1})(u_{m}-u_{b}),
\end{split}
\normalsize
 \label{FKmodel_twoLayers}
\end{equation}
where $I_{ion}^{m}$ and $I_{ion}^{b}$ are the ion currents at the middle slice and at the boundaries next to the bulk, respectively. We numerically integrate Eq.~(\ref{FKmodel_twoLayers}) using as initial condition a rigidly rotating spiral for $u_{m}$  and a 
non-excited state for $u_{b}$. 
The results are in perfect agreement with the full model Eq.~(\ref{FKmodel}) in terms of $\omega$, $\Delta\tau_{d}^{c}$ (Fig.~\ref{Fig04}A), and even core size of the scroll wave at the middle slice $R_{m}$ (Section IX and Fig.~S6 in the SM \cite{SupMat}). For $l>d_{z}$, it is still possible to reduce the bulk into a single slice, but this strategy cannot be extended to the boundary layer. However, one can approximate the correction in the 2-slice model for larger boundary layers (Section X in the SM \cite{SupMat}), and find again an excellent agreement for $l=2d_{z}$ (Fig.~\ref{Fig04}B).

In the feedback regime, the bulk, represented by $u_{m}$, experiences not only a decrease in excitability caused by $-2D_{z}\chi_{m+1}u_{m}/d_{z}^{2}$, but also a continuous forcing from the boundaries given by $2D_{z}\chi_{m+1}u_{b}/d_{z}^{2}$. To shed light on the nature of this forcing, close to $\alpha_{f}$, we can simplify the equation of $u_b$ in~(\ref{FKmodel_twoLayers}) by assuming that its dynamics is mainly driven by $u_{m}$; $\partial_{t}u_{b}\approx\alpha D_{z}(1-\chi_{b+1})(u_{m}-u_{b})/d_{z}^{2}$. This expression can be integrated to find $u_{b}$, which, when introduced into the first equation of~(\ref{FKmodel_twoLayers}), gives 
\begin{equation}
\partial_{t}u_{m}\approx D_{o}\nabla_{\perp}^{2}u_{m}
   - I_{\rm ion}^{m}
   - \frac{1}{\mathcal{T}_{b}}\,u_{m}
+ \frac{1}{\mathcal{T}_{b}\,\mathcal{T}_{bl}}
     \int_{0}^{t}\!e^{-\frac{t - t'}{\mathcal{T}_{bl}}}
         u'_{m}dt',
\label{forced_model}
\end{equation}
where $\mathcal{T}_{b}^{-1}\!\!=\!\!2D_{z}\chi_{m+1}/d_{z}^{2}$, $\mathcal{T}_{bl}^{-1}\!\!=\!\!\alpha D_{z}(1-\chi_{b+1})/d_{z}^{2}$ and $u'_{m}\!\!\!=\!\!\!u_{m}(x,y,t')$.
Integration of this forced model reproduces reasonably well the enhancement of stability (Fig.~\ref{Fig04}A).
Thus, this simplification reveals that wave stabilization is the consequence of a reduction in bulk excitability and a boundary-driven delayed-feedback. A useful lesson of Eq.~(\ref{forced_model}) is that to sustain wave dynamics the feedback from the boundaries needs to be fast enough to balance the loss in bulk excitability, i.e., $\mathcal{T}_{b}/\mathcal{T}_{bl}=h/2d_{z}>1$. 
This last condition is not satisfied for the case $h\!=\!2d_{z}$, where we observe wave termination when  $l\!=\!d_{z}$ (Fig.~\ref{Fig02}B), indicating that the bulk cannot support waves due to a loss in excitability. The forced model also captures  $\Delta\tau_{d}^{c}$ when $l=2d_z$ (Fig.~\ref{Fig04}B), and reveals that larger boundary layers correlate with smaller $\mathcal{T}_{b}/\mathcal{T}_{bl}$ (Section X and Fig.~S7 in the SM \cite{SupMat}).

The stabilizing properties of boundary layer heterogeneities are not limited to the meandering instability but can be extended to spiral wave break-up. To illustrate this, we modify the electrophysiological parameters (SM \cite{SupMat}; Section I) in the FK model such that a rigidly rotating spiral wave in homogeneous conditions undergoes break-up driven by discordant alternans \cite{fenton2002multiple}. We  use the computationally efficient Eq.~(\ref{FKmodel_twoLayers}) to implement the boundary layer effects, but we have verified that the full \searr{quasi-2D} equations give
similar results. The simulations reveal that the introduction of boundary layers can enhance the stability of scroll waves and prevents break-up initiation (Section XI and Fig.~S8 in the SM \cite{SupMat}).

In summary, we prove that boundary layer heterogeneities can control scroll wave dynamics through either heterogeneities in the coupling strength or the tissue excitability (SM \cite{SupMat}, Section XII). Unlike prior studies that emphasize destabilizing gradients along the filament orientation \cite{storb2003tomographic,qiao2008control,biktashev2011evolution}, our contribution focuses on the stabilizing role and reveals theoretically the stabilization mechanism. It results from a balance between reduction in bulk excitability and a boundary-driven delayed-feedback: the boundary serves as a slowing zone that receives the incoming bulk wave and feeds it back with a delay that depends on geometry ($d_z,h,l$) and \searr{coupling strength} ($D_{z},\alpha D_{z}$). \searr{While feedback-mediated stabilization of spiral waves has been studied before using imposed measurements and controllers \cite{schlesner2006stabilization}, in our system the feedback emerges intrinsically from a bulk-boundary layer interaction.} Additionally, we identify a leakage-feedback transition that resembles a pinning-unpinning transition. \sea{This underscores} how the semidiscrete limit, physiologically plausible in thin heterogeneous cardiac tissue, uncovers novel dynamical regimes in excitable media.

Our work could pave the way for new control strategies in cardiac research, where the tuning of boundary heterogeneities could enable bulk waves to self-regulate themselves. Of course, it will be necessary to characterize tissue properties to avoid scenarios in which the partially ablated region could favor wave break-up of incoming planar waves, a common ablation caveat  \cite{anter2015high,gottlieb2021blinding}. Cardiac optogenetics \cite{hussaini2024efficient,li2023reordering,nyns2022optical,hussaini2021drift,bruegmann2016optogenetic} may provide a way to
precisely manipulate boundary  properties and experimentally test our mechanism. Future work could investigate the long-term effects of  ablation, where healing processes and nonlocal mechanical effects may play a relevant role. \searr{Finally, we expect boundary-driven control to be applicable to other excitable systems, such as Rho GTPase wave activity in the actin cortex of eukaryotic cells \cite{bement2015activator} and the Belousov-Zhabotinsky chemical reaction \cite{zaikin1970concentration}, motivating further experiments.}
 
We thank Yuhai Tu, Mahesh Kumar Mulimani and Michael Reiss for fruitful discussions. This work was financially supported by NIH R01 HL122384. S.E.-A. acknowledges the financial support of Beca Chile 74230063.

The data that support the findings of this article are openly available \cite{echeverria_alar_2025_codes}.

% Bibliography
%\bibliography{apssamp.bib}

\begin{thebibliography}{85}
\bibitem [{Acu()}]{AcuteNote}%
  \BibitemOpen
  \href@noop {} {}\bibinfo {note} {The timescales addressed in our analysis are on the scale of seconds (<4) after the heterogeneities are applied.}\BibitemShut {Stop}%
\bibitem [{bif()}]{bifurcationNote}%
  \BibitemOpen
  \href@noop {} {}\bibinfo {note} {This bifurcation occurs for exactly the same value in the 2D version of Eq.~(\ref{FKmodel}) (i.e., $\mathcal{C}_{z}=0$), and is independent of the thickness $\mathcal{H}$, at least when $\mathcal{H}\leq 34d_{z}$.}\BibitemShut {Stop}%
\bibitem [{log()}]{logNote}%
  \BibitemOpen
  \href@noop {} {}\bibinfo {note} {The $\alpha$ values used in this study are nearly exponentially distributed within the range $[2\!\times\!10^{-5},2\!\times\!10^{-1}]$. The logarithmic scale in Figs.~\ref{Fig02}C-~\ref{Fig02}D is introduced to plot the phase diagrams on a linear scale.}\BibitemShut {Stop}%
\bibitem [{HHn()}]{HHnote}%
  \BibitemOpen
  \href@noop {} {}\bibinfo {note} {The maximum slab thickness systematically analyzed is $34d_z=850$ $\mu$m. This is not a limiting value but rather a choice driven by the numerical cost of studying the discrete limit along $z$. For example, the stabilization mechanism is still present in a slab of size $70d_z=1.75$ mm, using boundary layer heterogeneities of size $l=5d_z$ and strength $\alpha=0.08$. The enhancement of stability is $\Delta\tau_{d}^{c}/\tau_{d}^{c}=0.029$, which is equivalent to the case $\mathcal{H}=14d_z$ and $\alpha=0.004$ (Fig.~\ref{Fig02}A)}\BibitemShut {NoStop}%
\bibitem [{con()}]{continuumNote}%
  \BibitemOpen
  \href@noop {} {}\bibinfo {note} {Notice that we have introduced the continuum limit along $z$, which is valid in the bulk.}\BibitemShut {Stop}%
\end{thebibliography}
\providecommand{\noopsort}[1]{}\providecommand{\singleletter}[1]{#1}%

\end{document}

\preprint{APS/123-QED}

\title{Supplemental Material for Boundary-driven delayed-feedback control of spatiotemporal dynamics in excitable media}

\author{Sebasti\'an Echeverr\'ia-Alar}
\author{Wouter-Jan Rappel}%

\affiliation{Department of Physics, University of California, San Diego, California, 92093, USA}

\date{\today}% It is always \today, today,

\maketitle

\setcounter{figure}{0}
\setcounter{table}{0}

\tableofcontents
\newpage

\section{Electrophysiological model}
The reaction-diffusion model chosen to explore the effects of boundary layer heterogeneities in the stability of scroll waves, is a semi-discrete version of the well-established Fenton-Karma (FK) model \cite{fenton1998vortex}.
In this model, the evolution of the transmembrane action potential $u$ obeys
\begin{equation}
\partial_{t}u = \mathcal{I}_{k}D_{o}\nabla_{\perp}^{2}u + \frac{D_z}{d_z^2}C_{z} - \dfrac{I_{ion}}{C_{m}},
\label{Eq1}
\end{equation}
where $D_{o}=0.001$ cm$^{2}$/ms is a diffusion coefficient and $\mathcal{I}_{k}$ is a function controlling the heterogeneity distribution in each slice:
\begin{equation}
    \mathcal{I}_{k}=
    \begin{cases}
      \alpha , & 0\leq kd_{z} < l \cup \mathcal{H}-l< kd_{z} \leq \mathcal{H}  \\
      1, & l \leq kd_{z} \leq \mathcal{H}-l
    \end{cases}.
    \label{alpha}
  \end{equation}
Here, the subscript $k=\{0,..,\mathcal{H}/d_z\}$ labels the tissue slice along the transmural direction. The second term, $C_{z}$, describes discrete coupling along the $z$ direction and is given by
\begin{equation}
 \mathcal{C}_{z} =  \Bigl[\mathcal{Q}_{k+1}(u_{k+1} - u_{k}) + \mathcal{Q}_{k-1}(u_{k-1}-u_{k})\Bigr], 
 \label{zCoupling}
\end{equation}
where $\mathcal{Q}_{k+1}=min(\mathcal{I}_{k},\mathcal{I}_{k+1})$ and $\mathcal{Q}_{k-1}=min(\mathcal{I}_{k},\mathcal{I}_{k-1})$. Consistent with experimental data, we use a transverse diffusion constant $D_{z}=D_{o}/5$. $C_m$ is the membrane capacitance and $I_{ion}=I_{fi}+I_{so}+I_{si}$ is the sum of ionic currents. The membrane capacitance is dimensionless and equal to 1 and has thus been omitted in the main text. The currents depend on electrophysiological parameters
and describe the following excitable dynamics
\begin{equation}
I_{fi}(u,v) = -\dfrac{v}{\tau_{d}}\Theta(u-u_{c})(1-u)(u-u_{c}),
\label{Eq2}
\end{equation}

\begin{equation}
I_{so}(u) = \dfrac{u}{\tau_{o}}\Theta(u_{c}-u) + \dfrac{1}{\tau_{r}}\Theta(u-u_{c}),
\label{Eq3}
\end{equation}

\begin{equation}
I_{si}(u,w) = -\dfrac{w}{\tau_{si}}(1 + \tanh(k(u-u_{c,si}))).
\label{Eq4}
\end{equation}

The gating variables $v$ and $w$ evolve according to local dynamics:
\begin{equation}
\partial_{t}v = \dfrac{1}{\tau_{v}^{-}(u)}\Theta(u_{c} - u)(1-v) - \dfrac{1}{\tau_{v}^{+}}\Theta(u - u_{c})v ,
\label{Eq5}
\end{equation}

\begin{equation}
\partial_{t}w = \dfrac{1}{\tau_{w}^{-}}\Theta(u_{c} - u)(1-w) - \dfrac{1}{\tau_{w}^{+}}\Theta(u - u_{c})w ,
\label{Eq6}
\end{equation}
where $\tau_{v}^{-}(u)=\Theta(u - u_{v})\tau_{v1}^{-} + \Theta(u_{v} - u)\tau_{v2}^{-}  $. In our simulations, all electrophysiological parameters are 
kept constant (and listed in  Table~\ref{TabS1}), except   $\tau_{d}$, which controls the excitability in Eq.~(\ref{Eq1}), and is used as the bifurcation parameter to trigger either a meandering or a break-up instability. The system of equations is numerically integrated on   
a slab of size $L\times L \times \mathcal{H}$ (see Fig.1 in the main text).
In each $x-y$ slice, the Laplacian is approximated by a 5-point stencil with a spatial discretization size $\Delta x=0.0254$ cm, and in the transmural ($z$) direction, we used the discretization reported in the main text. We have verified that the effects of choosing smaller values of $\Delta x$ are negligible (see Section XIII). \searr{The whole slab is bounded by non-flux boundary conditions.} We used
 a forward Euler method for temporal integration with a time step $\Delta t=0.01$ ms for the quasi-2D simulations, $\Delta t=0.08$ ms for the reduced 2D simulations, and $\Delta t = 0.005$ ms for the simulations in Section XIII. \sea{The numerical implementation was done in Matlab and WebGL \cite{echeverria_alar_2025_codes}. The latter is briefly detailed in Section XIV.}
\begin{table}[htbp]
\centering
\caption{Electrophysiological and numerical parameters used for all numerical simulations in this study. The first (second) row corresponds to the set of parameters for which a spiral wave, at a critical $\tau_{d}$ value, undergoes a meandering (break-up) instability in homogeneous conditions.}
\label{TabS1}
%\begin{center}
%\setlength{\tabcolsep}{20pt}
%($\mu$F/cm$^{2}$)
\addtolength{\tabcolsep}{0pt}
\begin{tabular}{@{}l c c c c c c c c c c c c c c c c}\hline
\hspace{0.45cm}Instability & $C_{m}$ & $u_{c}$ & $\tau_{o}$ & $\tau_{r}$ & $\tau_{si}$ & $k$ &  $u_{c,si}$ & $u_{v}$ & $\tau_{v1}^{-}$ & $\tau_{v2}^{-}$ & $\tau_{v}^{+}$ & $\tau_{w}^{-}$ & $\tau_{w}^{+}$ & $L$ & $\Delta x$ & $d_{z}$ \\
 & (-) & (-) & (ms) & (ms) & (ms) & (-) & (-) & (-) & (ms) & (ms) & (ms) & (ms) & (ms) & (cm) & (mm) & ($\mu$m) \\
\hline
\rule{0pt}{1.0\normalbaselineskip}
\hspace{0.5cm}Meandering & $1$ & $0.13$ & $9$ & $33$ & $29$      & $15$ & $0.5$ & $0.04$ & $9$ & $8$ & $3.3$ & $60$ & $250$ & $5.08$ & $0.254$ & 25 \\
\hline
\rule{0pt}{1.0\normalbaselineskip}
\hspace{0.5cm}Break-up & $1$ & $0.15$ & $9$ & $34$ & $26.5$  & $15$ & $0.45$ & $0.04$ & $15.6$ & $5$ & $3.3$ & $80$ & $350$ & $12.24$ & $0.254$ & 25 \\
\hline
\end{tabular}
\addtolength{\tabcolsep}{-3pt}
%\end{center}
\end{table}

\section{Scroll wave stabilization in the presence of a localized transmural heterogeneity} 
\begin{figure}[htbp]
\includegraphics [width=14.5 cm] {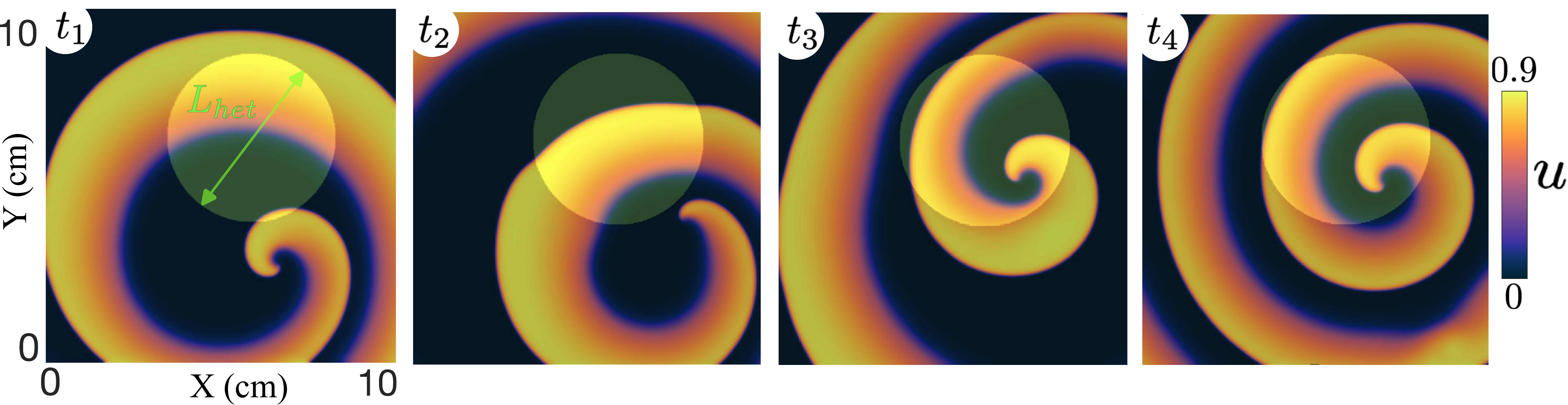} 
\caption{Scroll wave interacting with a localized boundary layer heterogeneity in the FK model. The temporal sequence $t_{1}= 0 < t_{2} <t_{3}< t_{4}=19T_{rot}$, with $T_{rot}=122$ ms, describes the trapping of a scroll wave observed at the middle of a tissue slice with thickness $\mathcal{H}=14d_{z}$. The boundary layers have a volume $\pi L_{het}^{2}l/4$ and in this particular simulation $l=d_{z}$ and $L_{het}=4$ cm.}
\label{FigS1}
\end{figure}

In the main text, we focus on the simple scenario in which the heterogeneity distributions are uniform in the $x-y$ directions and encompass the whole slice. 
Here, we ask ourselves if it is possible to stabilize the meandering instability with a localized heterogeneity in the $x-y$ plane and if 
this heterogeneity can trap a meandering scroll wave. To answer this question, we introduce a circular heterogeneity 
in the upper and lower slice of a $2L\times2L\times\mathcal{H}$ slab. The diameter of the 
heterogeneity is taken as $L_{het}=4$ cm and the reduction of 
conduction is parameterized by $\alpha=0.004$. Note that this reduction now occurs not only in the transmural direction but also in the $x-y$ plane.
Fig.~\ref{FigS1} illustrates snapshots of a spiral wave that is 
meandering in the absence of a boundary layer heterogeneity ($\tau_{d}=0.382$). The introduction of the circular  
heterogeneity, however,  stabilizes and traps the meandering wave: once the filament enters the heterogeneous region (circular shaded region), it starts to exhibit locally a rigid rotation behavior ($t_{3}$) and after a transient of around $18T_{rot}$, the final equilibrium state is a rigidly rotating scroll wave ($t_{4}$).
%7.5

\section{Transmural rotational anisotropy}
Cardiac tissue is inherently anisotropic \cite{rappel2022physics}. The elongated cardiomyocytes transport ions more efficiently along their longitudinal axis than along their 
transverse axis. In 2D simulations, this anisotropy is usually not an issue, as differences in conductivity along the $x$ and $y$ directions can be rescaled, as long as the orientation of the cardiac cells does not vary in space. In 3D, however, the situation becomes more subtle: even if the orientation of the cells is independent of $x$ and $y$, it can change along the $z$-axis. 
Motivated by  images of the myocardial architecture in the ventricles, the resulting transmural dependence is generally modeled in simulations as a constant change in cardiomyocytes orientation (rotational anisotropy) from endocardium to epicardium \cite{fenton1998vortex}. Simulations that incorporated
rotational anisotropy have revealed that it can only de-stabilize a spiral wave filament if 
the tissue thickness is sufficiently large \cite{fenton1998vortex,rappel2001filament}. In other words, in thin geometries such as the 
atrial walls, rotational anisotropy will only have a negligible effect in scroll wave dynamics.

We have verified that in our thickest tissue slab ($\mathcal{H}=34d_{z}$), even using a strong rotational rate of $30^{\circ}$/mm \cite{fenton1998vortex,rappel2001filament}, transmural rotational anisotropy does not modify the critical $\tau_{d}$ values reported in the main text, neither in the homogeneous case ($\tau_{d}^{c}$) nor in the heterogeneous ($\tau_{d}^{c*}$) case with $l=d_{z}$ and $\alpha=0.004$.
Thus, it is reasonable to neglect transmural anisotropy.

\section{Boundary layers in thin curved geometries}
\begin{figure}[htbp]
\includegraphics [width=13 cm] {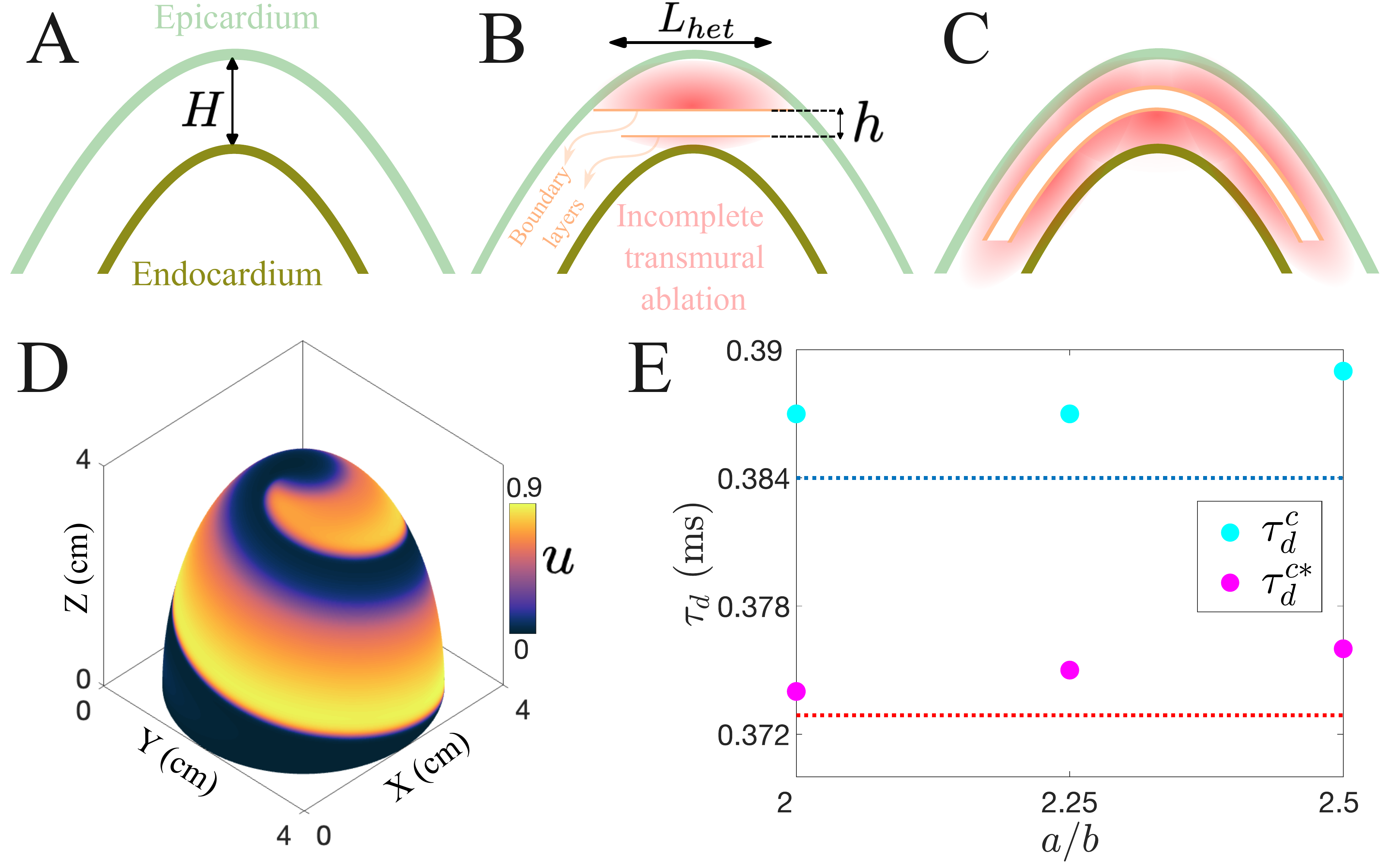} 
\caption{Boundary layer effect on the meandering instability in a thin half-ellipsoidal geometry. (A) Two-dimensional cross-sectional cut of the curved geometry. (B-C) Schematic representations of  incomplete transmural ablation lesions, represented in red, in the curved geometry. 
In B, the lesion is small 
compared to the inverse of the curvature and the system can be treated as in the main text. In C, the size of the lesion cannot be 
neglected compared to the curved geometry. (D) Initial transmembrane voltage at the outer (epicardial) surface of the half-ellipsoid. (E) Critical $\tau_{d}$ values for the homogeneous ($\tau_{d}^{c}$) and the heterogeneous cases ($\tau_{d}^{c*}$) with $\alpha=0.004$. The dashed lines correspond to the critical values in the non-curved geometry (see main text).}
\label{FigCURVED}
\end{figure}

To show the stabilization of scroll waves against meandering instabilities in more complex geometries, we consider a half-ellipsoidal shape with a nearly constant thickness $\mathcal{H}$ (a 
cross-sectional cut is shown in Fig.~\ref{FigCURVED}A). This geometrical representation has been used before to study scroll wave dynamics based on morphological measurements of the human heart \cite{pravdin2015drift}. The introduction of the boundary layers can be motivated by two possible incomplete ablation lesions. For example, if the ablation produces a gap, of size $h$, such that the curvature from the heart surfaces can be neglected compared to the spatial dimension of the gap, as shown in Fig.~\ref{FigCURVED}B, our theoretical framework remains valid given that $L_{het}$ is bigger than the spiral core (see Fig.~\ref{FigS1}).  A less trivial case is presented in Fig.~\ref{FigCURVED}C, where,
possibly due to multiple unfinished ablations with different orientations, the gap displays a distinct curvature.

We use the phase field approach to model the case depicted in Fig.~\ref{FigCURVED}C, where non-flux boundary conditions on the inner and outer surfaces of the thin half-ellipsoidal geometry are enforced via a scalar field $\varphi$ \cite{fenton2005modeling}. This quantity, defined on a Cartesian grid, is equal to $1$ inside the half-ellipsoid and is $0$ outside. The interface connecting both values is  generated through a diffusion process, using $\phi=1$ in the region of interest and $0$ elsewhere as initial condition. By varying the total evolution time of the diffusive dynamics, one can control the smoothness (or sharpness) of the interface. We implement the phase field so that any one-dimensional cut in the transmural direction closely matches $\mathcal{I}_{k}$  (see Eq.~2 in the main text) with $\mathcal{H}=14d_{z}$ and $l=d_{z}$. For the homogeneous scenario, i.e., without a boundary layer, we apply the Closest Point Method, which allows to impose non-flux boundary conditions on curved surfaces (see \cite{macdonald2013simple} for details). Notice that, because our geometries are thin, the quasi-2D homogeneous scenario is equivalent to the 2D one.
Once the geometries are generated, we integrate the FK model using the parameters for 
which the spiral wave undergoes a meandering instability (first row of Table~\ref{TabS1}).

The initial condition is a scroll wave placed on top of a half-ellipsoid. The epicardial transmembrane voltage is shown in 
Fig.~\ref{FigCURVED}D for a major axis $a=4$ cm and a minor axis $b=2$ cm. Then, as in the main text, we decrease  $\tau_{d}$ from $\tau_{d}^{o}=0.390$ until  a meandering instability is triggered. 
In the homogeneous case, the critical value of $\tau_{d}$, $\tau_{d}^{c}$, is increased 
compared to the non-curved geometry (cyan symbols vs. the blue dashed line in Fig.~\ref{FigCURVED}E).
In the presence of boundary layers the meandering stability is enhanced,  $\tau_{d}^{c*}<\tau_{d}^{c}$,
(magenta symbols, Fig.~\ref{FigCURVED}E) with a slight increase in $\tau_{d}^{c*}$ compared to the 
non-curved case (red dashed line). This enhancement is robust against variations of the 
curvature, controlled by the ratio  $a/b$.

\section{Larger boundary layers and wave termination}
\begin{figure}[htbp]
\includegraphics [width=12 cm] {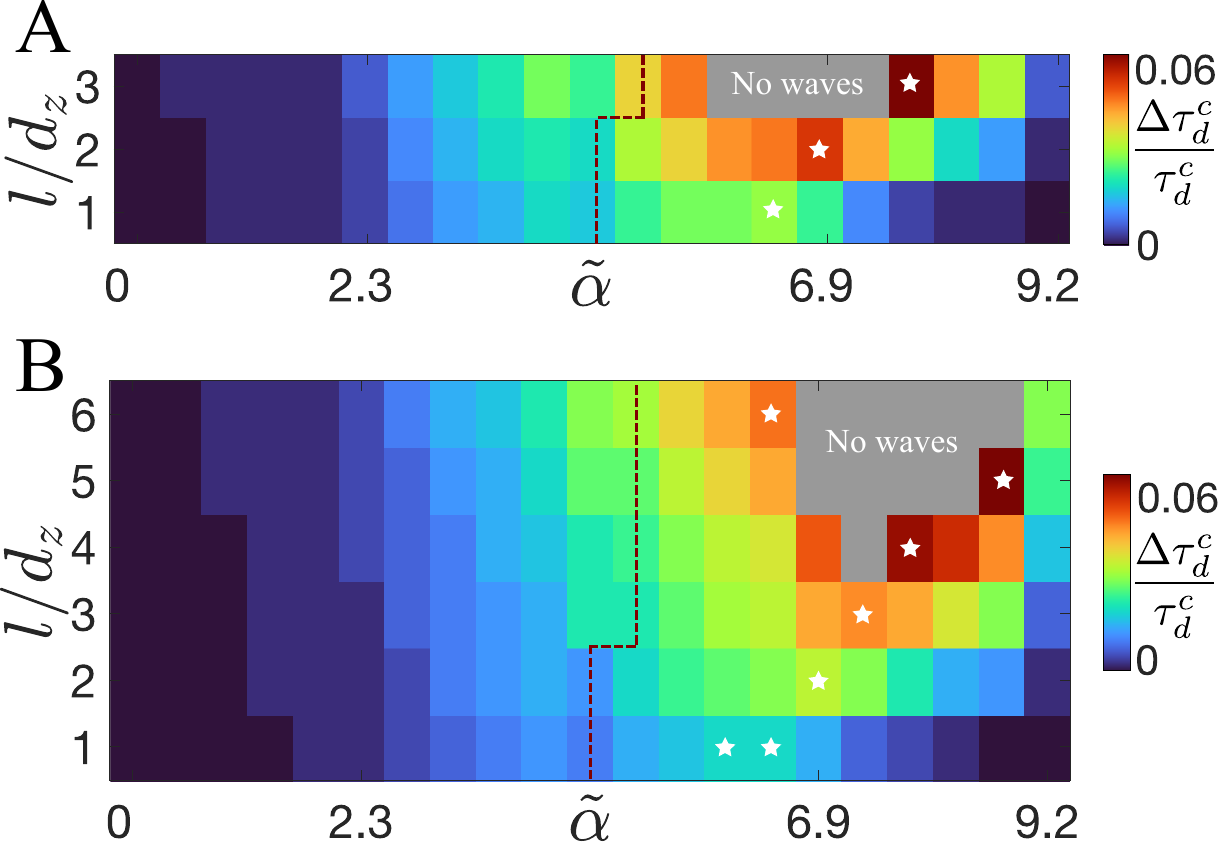} 
\caption{Effects of boundary layer size $l$ in the enhancement of scroll wave stability $\Delta\tau_{d}^{c}/\tau_{d}^{c}$ for different $\alpha$ values, where $\tilde{\alpha}=log(\alpha)+10.82$, and for two slab thicknesses: (A) $\mathcal{H}=14d_{z}$ and (B) $\mathcal{H}=24d_{z}$. The dashed red lines indicate the propagation failure along the transmural direction. Gray regions: termination of all wave activity. \searr{The white stars indicate $\max(\Delta\tau_{d}^{c}/\tau_{d}^{c})$ for each boundary layer size}.}
%\vspace{-18mm}
\label{FigS2}
\end{figure}

In the main text, we focus on the scroll wave stabilization using boundary layer heterogeneities of size $l=d_{z}$. However, we have verified that the delay of the meandering instability by slowing down scroll waves is also present when $l>d_{z}$. Fig.~\ref{FigS2} shows the change in the enhancement of stability of scroll waves, 
$\Delta\tau_{d}^{c}/\tau_{d}^{c}$,
in tissue slab of thicknesses $\mathcal{H}=14d_{z}$ (A) and $\mathcal{H}=24d_{z}$ (B) in the presence of  boundary layers of sizes $l<\mathcal{H}/4$. This choice implies that we only consider situations where the bulk is bigger than the boundary layers; $h>2l$. When $l>2d_{z}$, our simulations show that the leakage-transition feedback is shifted towards higher values of $\alpha$, 
as indicated by the dashed line in Fig.~\ref{FigS2}.

In the feedback regime ($\alpha\geq\alpha_{f}$), the enhancement of stability is larger for thicker boundary layers. This is  related to the fact that the ratio $2l/h$ becomes closer to $1$. Additionally, we observe that in thicker boundary layers, a significant increase in $\Delta\tau_{d}^{c}/\tau_{d}^{c}$ can be achieved for higher $\alpha$ values. 

As in the main text, the gray zones indicate regions in which wave activity is terminated. 
For the larger slab thicknesses, termination occurs over a range of boundary layer 
sizes. Furthermore, this termination also shifts towards larger values of $\alpha$. Thus,  
 smaller differences in conductivity between the bulk and the boundary layer will be able to terminate all wave activity.

 \section{Asymmetric heterogeneity distribution}
\begin{figure}[htbp]
%17.5
\includegraphics [width=17. cm]{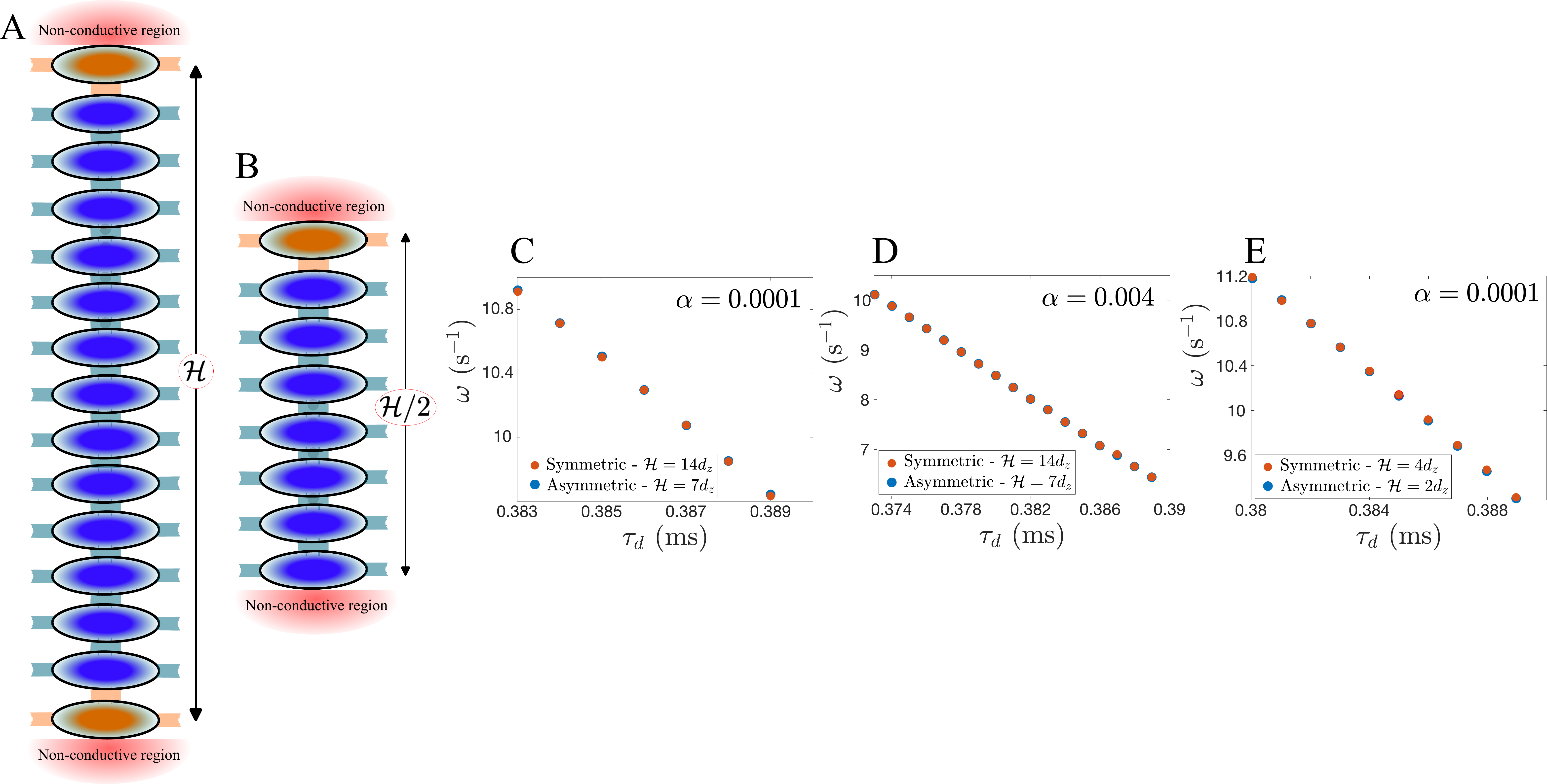} 
\caption{Comparison between symmetric and asymmetric heterogeneity distributions. (A-B) Schematic representation of the coupling along the transmural direction (see Fig. 1 in the main text for further detail). (C-E) Angular frequency as a function of the excitability parameter $\tau_{d}$ for different values of $\alpha$ and $\mathcal{H}$.}
\label{FigS3half}
\end{figure}

The symmetry of the heterogeneity distribution used in the main text (see Fig.~\ref{FigS3half}A) allows us to easily extend our results to asymmetric distributions in thin tissue slabs. Fig.~\ref{FigS3half}B displays an asymmetric distribution consisting of a tissue slab of half the size in the symmetric case and with a heterogeneous boundary layer in only one boundary. This new numerical setup, as in the symmetric case, is bounded by non-flux boundary conditions. Similar to the analysis in the main text, we sweep the excitability parameter $\tau_{d}$ starting from a rigidly rotating scroll wave and found that the angular frequency and the enhancement of stability are equal when comparing the symmetric and asymmetric cases for different values of $\alpha$ and $\mathcal{H}$ (Figs.~\ref{FigS3half}C-E). We notice that in both $\mathcal{H}=4d_{z}$ (symmetric) and $\mathcal{H}=2d_{z}$ (asymmetric) cases, scroll waves terminate in the presence of a heterogeneity with strength $\alpha=0.004$.

%\section{Weakly nonlinear analysis in the leakage regime \scalebox{1.3}
\section{Weakly nonlinear analysis in the leakage regime $\alpha<\alpha_{f}$}
Due to the symmetry of our problem, one can consider only  one half of the tissue slab. For example, in the case of a slab with three slices and with a heterogeneity of size $l=d_{z}$ and strength $\alpha$, we can write 6 2D equations

\begin{equation}
\begin{split}
\partial_{t}u_{1} &= D_{x}\nabla_{\perp}u_{1} + 2\alpha\dfrac{D_{z}}{d_{z}^{2}}(u_{0}-u_{1}) -\dfrac{{I}_{ion}}{C_{m}}\\
\partial_{t}v_{1} &= {F}(u_{1},v_{1})\\
\partial_{t}w_{1} &= {G}(u_{1},w_{1})\\
\partial_{t}u_{0} &= \alpha D_{x}\nabla_{\perp}u_{0} + \alpha\dfrac{D_{z}}{d_{z}^{2}}(u_{1}-u_{0}) -\dfrac{{I}_{ion}}{C_{m}}\\
\partial_{t}v_{o} &= {F}(u_{o},v_{o})\\
\partial_{t}w_{o} &= {G}(u_{o},w_{o}).\\
\end{split}
 \label{FKmodel_6}
\end{equation}
Here the index $0$ refers to the boundary layer and the index 
$1$ refers to the bulk.
In the leakage regime, $u_{1}\gg u_{0}$, we only need to take into account  equations in 
the bulk since $u_{0}-u_{1}\approx -u_{1}$. In other words,  the bulk receives
a negligible amount of current from the boundary.

Our analysis relies on the linearization of some of the nonlinear terms in Eq.~(\ref{FKmodel_6}). Therefore,  we need to smooth the Heaviside functions and 
we replace these functions by sigmoidal functions; $\Theta(q) = (1 + \tanh(s_{f}q))/2$ with $s_{f}=15$. This changes the excitability of the system and, therefore, the behavior and shape of spiral waves in the parameter set of Table~\ref{TabS1}. However, it is still possible to trigger a meandering instability, now at $\tau_{d}^{c}=0.348$ instead of $0.384$ as for the Fenton-Karma model considered in the main text.

We numerically solve the bulk part of Eq.~(\ref{FKmodel_6}) with $\alpha=0$ and smoothed Heavisides in polar coordinates where the initial condition is a spiral wave generated from direct numerical simulations in a cartesian grid, which is interpolated to a polar grid centered at the spiral core center. The polar geometry consist of a disk of radius $R_{disk}=2.29$ cm with a spatial discretization $\Delta r=\Delta x$ in the radial direction and $128$ points along the angular ($\theta$) direction. In the radial direction a 4-point stencil is used with non-flux boundary conditions at $r=0$ and $r=R_{disk}$, and a spectral method is used to discretize the angular terms of the Laplacian \cite{sandstede2023spiral}. The equations are integrated in time with an explicit Euler method with $\Delta t = 0.0001$ ms. The steady state, achieved after $15$ rotations of the spiral wave at $\tau_{d}=0.355$, is described by the spiral solution $\{u_{1},v_{1},w_{1}\}^{[0]}$ with an angular frequency $\omega_{o}$.

In the following, we perform a weakly nonlinear analysis around the homogeneous solution in the rotational frame of reference $\partial_{t}=\omega\partial_{\theta}$ to characterize how the frequency $\omega$ varies as a function of the heterogeneity strength $\alpha$. Introducing the ansatz $\mathbf{u}_{1}^{[0]}+\alpha\mathbf{u}_{1}^{[1]}+...$, $\omega=\omega_{o}+\Delta\omega+...$ in the bulk terms of Eq.~(\ref{FKmodel_6}) and considering that $\Delta\omega\sim\alpha$, we find at $\mathcal{O}(\alpha)$ a linear system of equations for the nonlinear correction; $\mathcal{L}\{u_{1},v_{1},w_{1}\}^{[1]}=\mathbf{b}$ where 
\begin{equation}
\mathcal{L}=
	\begin{pmatrix}
		\vspace{10pt}
		D_{x}\nabla^{2}_{\perp}+\omega_{o}\partial_{\theta}-\dfrac{\partial \hat{I}_{ion}}{\partial u_{1}} & \dfrac{\partial \hat{I}_{ion}}{\partial v_{1}} & \dfrac{\partial \hat{I}_{ion}}{\partial w_{1}}\\[\jot]
		\vspace{10pt}
		\dfrac{\partial \hat{F}}{\partial u_{1}} &\omega_{o}\partial_{\theta} + \dfrac{\partial \hat{F}}{\partial v_{1}} & 
  0\\[\jot]	
  \vspace{10pt}
  \dfrac{\partial \hat{G}}{\partial u_{1}} &
  0 &
  \omega_{o}\partial_{\theta} + \dfrac{\partial \hat{G}}{\partial w_{1}}
  \\[\jot]		
	\end{pmatrix}
\label{LinearOperator}
\end{equation}
and 

\begin{equation}
	%\resizebox{.9\hsize}{!}{$
			\mathbf{b}=-\Delta\omega\partial_{\theta}\begin{pmatrix}
		\vspace{5pt}
		u_{1}\\[\jot]
		\vspace{5pt}
		v_{1}\\[\jot]
		w_{1}		
	\end{pmatrix}^{[0]}+2\alpha\dfrac{D_{z}}{d_{z}^{2}}\begin{pmatrix}
		\vspace{5pt}
		u_{1}\\[\jot]
		\vspace{5pt}
		0\\[\jot]
		0		
	\end{pmatrix}^{[0]}.
\label{NoLineal}
\end{equation}
The terms $\hat{I}_{ion}$, $\hat{F}$ and $\hat{G}$ in Eq.~(\ref{LinearOperator}) depend on the smoothed version of the Heaviside functions. To solve the linear system, we must introduce an inner product to apply a solvability condition, i.e., 
the linear equation will have solution if and only if $\mathbf{b}$ is orthogonal to the $Ker\{\mathcal{L^{\dag}}\}$. Based on previous studies \cite{henry2002scroll}, we consider the inner product $\langle \mathbf{g}|\mathbf{h}\rangle=\int\int\mathbf{f}\cdot\mathbf{g}rdrd\theta$, and the elements of the kernel 
$Ker\{\mathcal{L^{\dag}}\}=\{\bar{u}_{1},\bar{v}_{1},\bar{w}_{1}\}$ are calculated numerically with a precision $\mathcal{O}(10^{-3})$. Finally, we apply the solvability condition $\langle\{\bar{u}_{1},\bar{v}_{1},\bar{w}_{1}\}|\mathbf{b}\rangle=0$ and obtain

\begin{equation}
\Delta\omega = 2\alpha\dfrac{D_{z}}{d_{z}^{2}}\dfrac{\langle\{\bar{u}_{1},\bar{v}_{1},\bar{w}_{1}\}|\{u_{1},0,0\}^{[0]}\rangle}{\langle\{\bar{u}_{1},\bar{v}_{1},\bar{w}_{1}\}|\partial_{\theta}\{u_{1},v_{1},w_{1}\}^{[0]}\rangle}=2\alpha\dfrac{D_{z}}{d_{z}^{2}}\dfrac{A}{B},
\label{solvability}
\end{equation}
with $A/B = -0.81$.

%\section{Transition at \scalebox{1.3}{$\alpha=\alpha_{f}$}}
\section{Pinning-unpinning-like transition at $\alpha=\alpha_{f}$}
\begin{figure}[htbp]
%7.5
\includegraphics [width=9 cm] {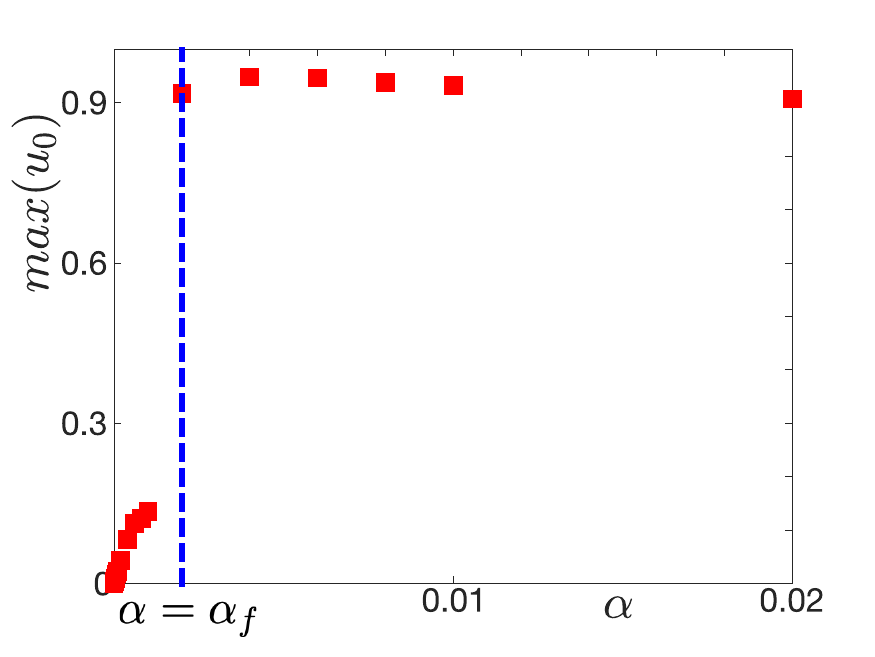} 
\caption{Maximum boundary voltage as a function of $\alpha$ in both leakage and feedback regimes for $\tau_{d}=0.390$, $\mathcal{H}=14d_{z}$ and $l=d_{z}$. 
}
\label{Fig3}
\end{figure}

For $\alpha<\alpha_{f}$, the boundary layer can not be excited by the bulk. As a consequence, the wave
is pinned to the boundary-bulk interface. 
In contrast, when $\alpha\geq\alpha_{f}$, wave propagation is possible from the bulk to the boundary and the wave is unpinned. This  sharp pinning-unpinning-like transition can be visualized by plotting the maximum value of the boundary transmembrane voltage $u_0$ as a function of $\alpha$, as shown in Fig.~\ref{Fig3}.

\section{Core radius in the multiple models}
\begin{figure}[h!]
%6.6
\includegraphics[width=9 cm]{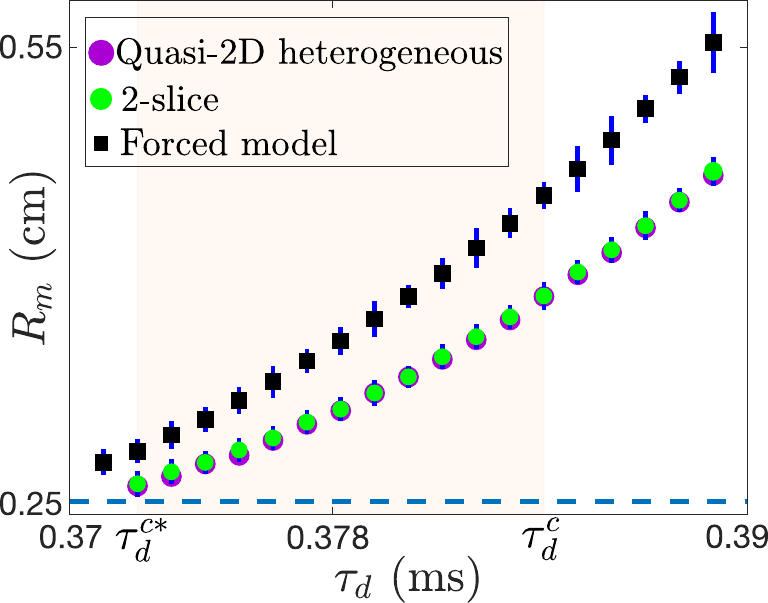}
\caption{Comparison of the core radius between the full heterogeneous model in a tissue slab of thickness $\mathcal{H}=14d_{z}$, with boundary heterogeneities of strength $\alpha=0.005$ and size $d_{z}$, the 2-slice model, and the forced model. Blue bars indicate standard deviation associated with errors in the tip detection algorithm. The dashed lines correspond to the critical $R_{m}$ value in the homogeneous case.}
%with $\chi_{1}=1/73$ and $\chi_{8}=1/2635$
\label{Fig04_core}
\end{figure}

The core radius $R_{m}$ of the scroll wave, measured in the middle slice of the thin slab, decreases as the $\tau_{d}$ parameter is also reduced (Fig.~\ref{Fig04_core}). This behavior is quantitatively reproduced by the 2-slice model and qualitatively emulated by the forced model. In the simulations of the forced model, Eq.~(3) in the main text, the integral is discretized using the trapezoidal rule, and the initial condition is a rigidly rotating spiral.

\section{Extension of the reduced models to larger boundary layers}
\searr{When $l>d_{z}$, it is still possible to reduce the bulk into a single slice by solving a Laplace equation. However, this technique cannot be applied to simplify the boundary layer, and a reduced model will need to take into account the
z-dependence of $\mathcal{C}_{z}$ in every boundary slice. This will result in a model with $1+l/d_{z}$ slices: the reduced bulk slice, the slice at the interface bulk-boundary layer, and the rest of $l/d_z-1$ slices within the boundary layer. Following the approach of the main text, a 3-slice model for the case $l=2d_z$ can be written as}

\begin{equation}
%\fontsize{9.4}{12}\selectfont
\begin{split}
\partial_{t}u_{m} &= D_{o}\nabla^{2}_{\perp}u_{m} - I_{ion}^{m} +2\dfrac{D_{z}}{d_{z}^{2}}\chi_{m+1}(u_{b}-u_{m})\\
\partial_{t}u_{b} &= \alpha D_{o}\nabla^{2}_{\perp}u_{b} - I_{ion}^{b} + \alpha\dfrac{D_{z}}{d_{z}^{2}}(1-\chi_{b+1})(u_{m}-u_{b})+\alpha\dfrac{D_{z}}{d_{z}^{2}}(u_{b-1}-u_{b})\\
\partial_{t}u_{b-1} &= \alpha D_{o}\nabla^{2}_{\perp}u_{b-1} - I_{ion}^{b-1} + \alpha\dfrac{D_{z}}{d_{z}^{2}}(u_{b}-u_{b-1}),
\end{split}
%\normalsize
 \label{FKmodel_3Layers}
\end{equation}
\searr{where the subscript $b-1$ refers to the transmembrane voltage and ionic currents in the extra boundary slice. Notice that the first equation in~(\ref{FKmodel_3Layers}) is equivalent to the case $l=d_z$, while the second equation has the extra term  $\alpha D_{z}(u_{b-1}-u_{b})/d_{z}^{2}$.}

\begin{figure}[htbp]
%14.6
\includegraphics[width=13 cm]{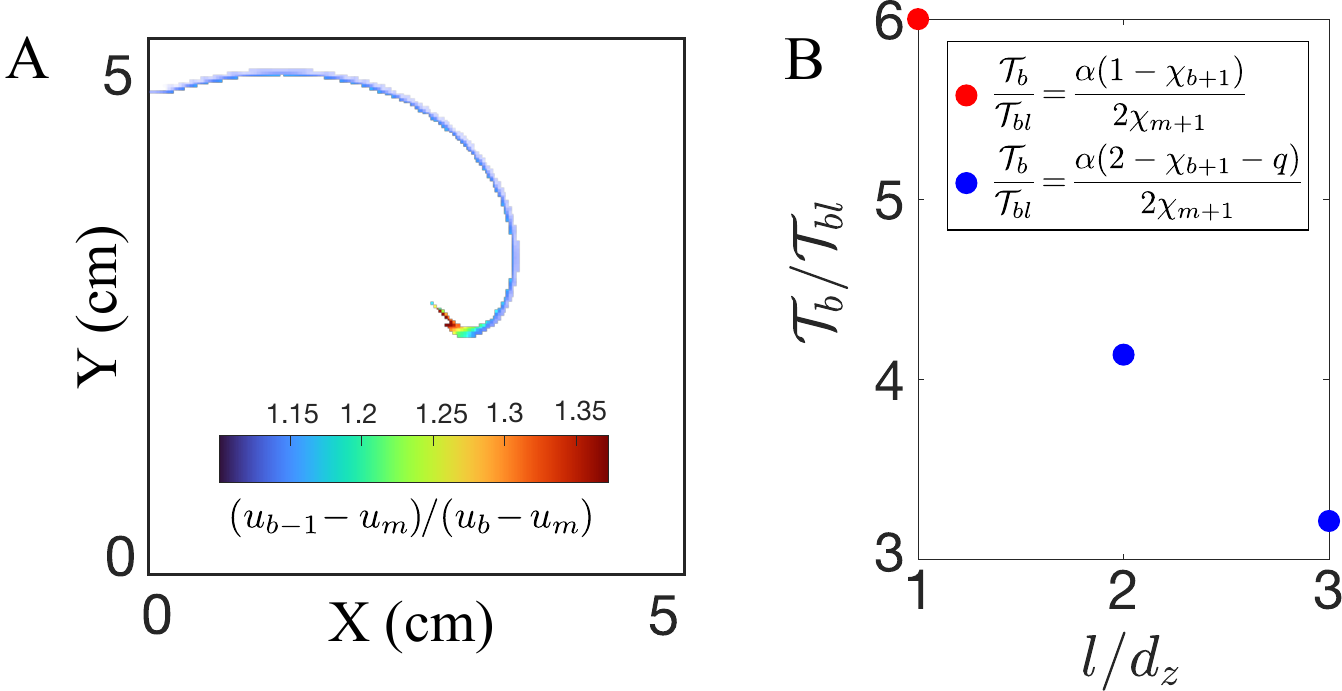}
\caption{\searr{Modeling larger boundary layer effects in a slab of thickness $\mathcal{H}=14d_{z}$ and heterogeneities of strength $\alpha=0.005$. (A) A temporal snapshot of the numerical ratio $(u_{b-1}-u_{m})/(u_{b}-u_{m})$ near the wave front. (B) The dependence of $\mathcal{T}_{b}/\mathcal{T}_{bl}$ on the size of the boundary layer. In the case $l=3d_z$, $q=1.19$.}}
\label{qPlot}
\end{figure}

\searr{Although Eq.~(\ref{FKmodel_3Layers}) is already compact, we ask ourselves if a 2-slice model can approximate the 3-slice model in order to gain mechanistic insights on how the size of the boundary layer affects the boundary-driven feedback. To address this question, it would be necessary to write $u_{b-1}$ as a function of $u_{b}$ and $u_{m}$. Inspired by our findings in the main text, we envision that a good approximation is $u_{b-1}=u_{m}+q(u_{b}-u_{m})$. We estimate $q$ by calculating the ratio $(u_{b-1}-u_{m})/(u_{b}-u_{m})$ from numerical integrations of the full quasi-2D equations~(\ref{Eq1}-\ref{Eq6}). A priori, $q$ is not a constant and it has $x-y$ dependency, which is a combination of numerical error from the model discretization and finite spatial gradients. We decide to only extract the ratio information close to the wave front of the spiral wave in the middle slice (Fig.~\ref{qPlot}A). This is motivated by the observation that the mismatch of transmembrane voltages at the bulk-boundary layer interface is stronger at the wave front (see the bottom panel of Fig.~3D in the main text). Finally, we take the spatial and temporal (over $15T_{rot}$) average of the numerical ratio, obtaining $q$. Then, in particular, we can rewrite the equation for $u_b$:}

\begin{equation}
\partial_{t}u_{b} \approx \alpha D_{o}\nabla^{2}_{\perp}u_{b} - I_{ion}^{b} + \alpha\dfrac{D_{z}}{d_{z}^{2}}(2-\chi_{b+1}-q)(u_{m}-u_{b}).
 \label{ub_qcase}
\end{equation}
\searr{Numerically integrating the above equation together with the first equation in~(\ref{FKmodel_3Layers}) yields the excellent agreement shown in Fig.~4B of the main text for $\mathcal{H}=14d_z$, $\alpha=0.005$, and $q=1.17$. This agreement indicates that $q$, computed only at $\tau_{d}^{o}$, depends solely on geometrical factors ($h$, $l$, $d_z$) and on $\alpha$, similar to $\chi_k$.}

\searr{The reduction from a 3-slice to a 2-slice model allows the derivation of a forced model as in the main text. In fact, this new forced model is equivalent to Eq.~(3) in the main text (corresponding to $l=d_z$), except that the time scale $\mathcal{T}_{bl}$ must be redefined as $\mathcal{T}_{bl}^{-1}=\alpha D_{z}(2-\chi_{b+1}-q)/d_{z}^{2}$. Note that $\chi_{b+1}$ and $\chi_{m+1}$, the latter controlling $\mathcal{T}_{b}$, are updated through their dependence on $h$ and $l$. Integration of the forced model reproduces reasonably well the enhancement of stability observed at $l=2d_z$ (Fig.~4B in the main text), which exceeds that for $l=d_z$. This increase in $\Delta\tau_{d}^{c}$ for larger boundary layers correlates with a decrease in the ratio $\mathcal{T}_{b}/\mathcal{T}_{bl}$ (Fig.~\ref{qPlot}B), indicating that thicker boundary layers produce slower boundary feedback relative to bulk excitability loss.}

\section{Control of spiral wave break-up through boundary layer heterogeneities}
\begin{figure}[htbp]
%14.6
\includegraphics[width=17 cm]{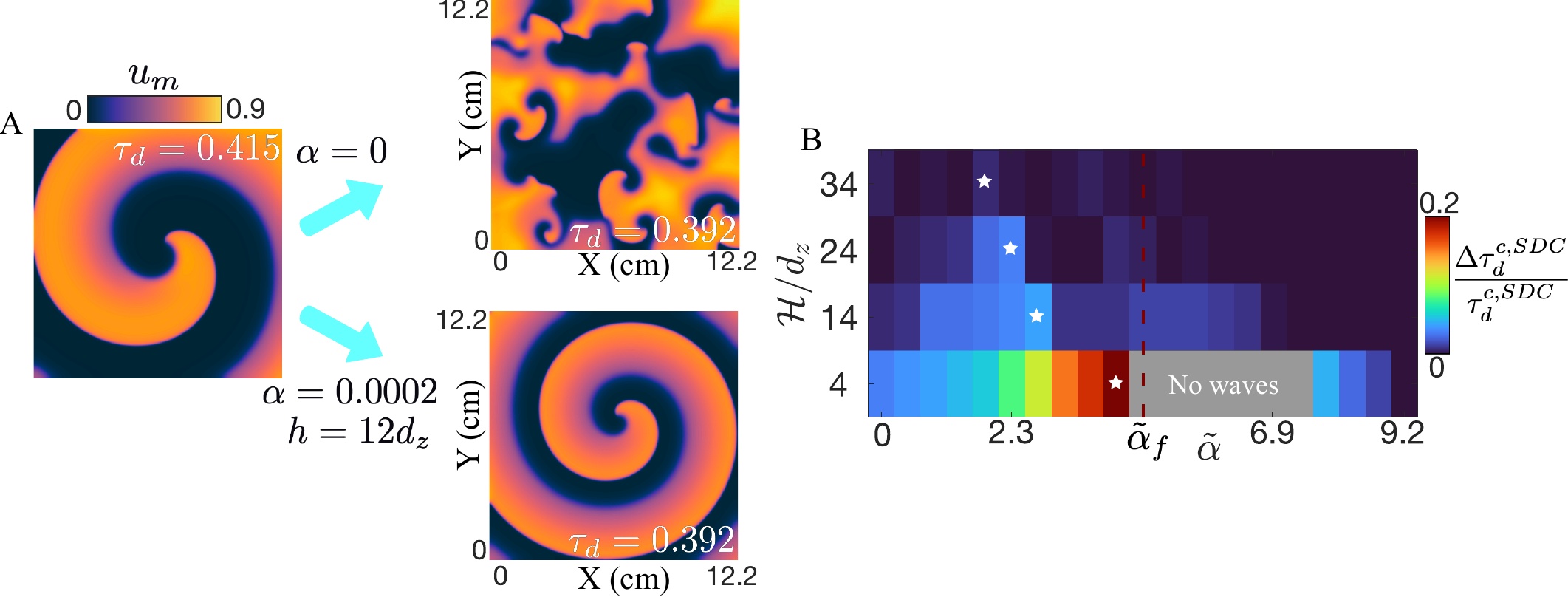}%{Fig5_02.pdf}
\caption{Control of spiral wave break-up through boundary layers of size $d_{z}$. (A) Spiral wave dynamics following a 
reduction of $\tau_{d}$ below the critical value in the homogeneous case (left panel) and in the presence of a boundary layer heterogeneity (right panel). (B) Phase diagram in $\tilde{\alpha}-\mathcal{H}/d_{z}$ space showing the level of enhancement of stability in the break-up case $\Delta\tau_{d}^{c,SDC}/\tau_{d}^{c,SDC}$. Gray region corresponds to parameter values for which wave 
activity is terminated. \searr{The white stars indicate $\max(\Delta\tau_{d}^{c}/\tau_{d}^{c})$ for each slab thickness.}}
\label{Fig05}
\end{figure}

The stabilizing properties of boundary layer heterogeneities are not limited to the meandering instability but can be extended to spiral wave break-up. To illustrate this, we modify the electrophysiological parameters (Table~\ref{TabS1}) in the FK model and use set 4 from Ref. \cite{fenton2002multiple} such that a rigidly rotating spiral wave in homogeneous tissue undergoes break-up. As detailed 
in this reference, this 
break-up is driven by discordant alternans when 
$\tau_{d}$ is smaller than a critical value $\tau_{d}^{c,SDC}=0.404$ ms (Fig.~\ref{Fig05}A). We use the computationally efficient Eq.~(2) from the main text, which offers a speed-up that can be estimated to be $(\mathcal{H}+d_{z})/(2d_z)$, to implement the boundary layer effects, but we have verified that the full quasi-2D equations give 
similar results. As in the main text, we start with a stable spiral wave and a parameter value larger than the critical value ($\tau_d=0.415$ ms) and decrease it in steps of $0.001$ ms every $30T_{rot}$. We use a bigger domain size, $L=12.24$ cm, in order to have sufficient space for the development of the break-up instability. The simulations reveal that the introduction of a boundary layer can enhance the stability of the spiral wave and prevents the break-up initiation. 
An example is shown in  Fig.~\ref{Fig05}A, 
which shows the stabilization in the case of $\alpha=0.0002$ and $h=12d_{z}$ ($\mathcal{H}=14d_{z}$). Instead of breaking up into multiple small-scale excitations, the spiral wave remains stable and displays a circular trajectory.

As in the main text, we perform a systematic variation of the thickness and the coupling strength, and construct a phase diagram showing the enhancement of stability in the break-up scenario $\Delta \tau_{d}^{c,SDC}/\tau_{d}^{c,SDC}$ (Fig.~\ref{Fig05}B). Interestingly, when $\mathcal{H}/d_{z}>4$, the strongest
 stabilization is observed in the leakage regime ($\alpha<\alpha_{f}$), different from the behavior in the meandering scenario. In this same regime, but when $\mathcal{H}/d_{z}\leq4$, we observe that spiral waves do not break up when sweeping down $\tau_{d}$; instead, they undergo meandering instabilities.

\section{Reduced excitability of boundary layer cells}
\begin{figure}[htbp]
\includegraphics [width=9. cm] {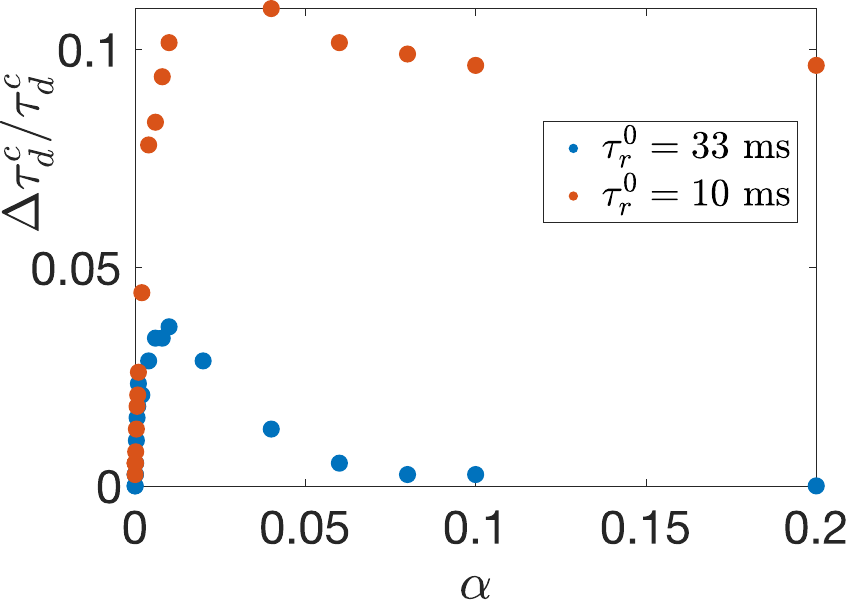} 
\caption{Enhancement of stability as a function of $\alpha$ in a tissue slab of thickness $\mathcal{H}=14d_{z}$ with a boundary layer of size $l=d_{z}$ and with  $\tau_{r}=33$ ms in the bulk. 
Plotted is the enhancement  in the presence of reduced excitability in the 
boundary layer ($\tau_{r}^{0}=10$ ms; red circles) and for the case of ionic homogeneity 
($\tau_{r}^{0}=33$ ms; blue circles). Note that for $\alpha>0.2$, the enhancement
saturates at $\Delta\tau_{d}^{c}/\tau_{d}^{c}=0.094$ in the presence of the ionic heterogeneity and 
that wave activity terminates between $\alpha=0.01$ and $\alpha=0.04$.
}
\label{hetIonic}
\end{figure}

In the main text, we model boundary layer heterogeneities using a reduction in the coupling strength of cardiomyocytes. An alternative and equally valid approach is to model boundary layer heterogeneities as a decrease in the excitability of cells. In our modeling framework, this can be easily implemented by decreasing $\tau_{r}^{0}$ in the boundary layer from its bulk value $\tau_{r}=33$ ms. We have verified that in this case the model still displays an enhancement of stability against the meandering instability, even for the case $\alpha=1$ (i.e., no heterogeneity in the coupling strength). For example, for $\tau_{r}^{0}=10$ ms, we find 
$\Delta\tau_{d}^{c}/\tau_{d}^{c}=0.094$ in a tissue slab of thickness $\mathcal{H}=14d_{z}$ and a boundary layer of size $l=d_{z}$. This enhancement is also present for $\alpha<1$, corresponding to a boundary heterogeneity in both the coupling strength and in the ionic current $I_{so}$ (red circles, Fig.~\ref{hetIonic}). Note that the enhancement of stability in this case is significantly increased when compared to the case addressed in the paper (blue circles, Fig.~\ref{hetIonic}). Obviously, this enhancement depends on the degree of excitability reduction, 
parameterized by $\tau_{r}^{0}$ \searr{(see below)}. Altogether, the equivalence in wave stabilization despite the type of heterogeneity can be qualitatively explained by the fact that boundary layer effects become important when the characteristic length of the wave is $\mathcal{O}(l)$. This length can be tuned by varying either the coupling strength or the time scales in the local dynamics of the transmembrane voltage.

\begin{figure}[htbp]
\includegraphics [width=10. cm] {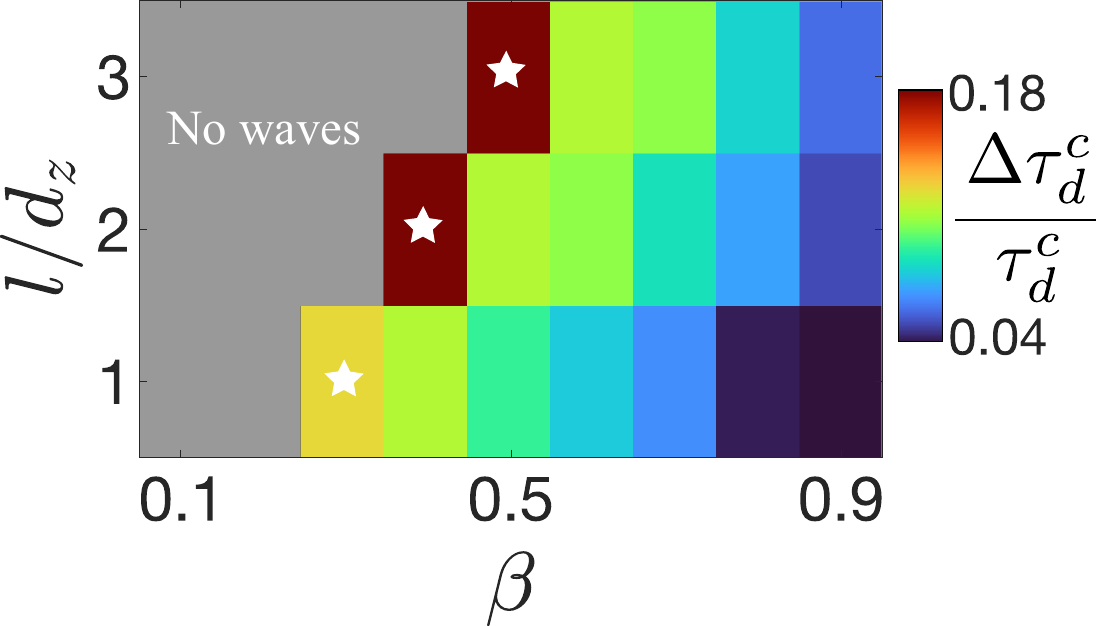} 
\caption{\searr{Effects of excitability reduction, parameterized by $\beta$, and boundary layer size $l$ in the enhancement of scroll wave stability $\Delta\tau_{d}^{c}/\tau_{d}^{c}$ in the wave break-up regime. The white stars indicate $\max(\Delta\tau_{d}^{c}/\tau_{d}^{c})$ for each boundary layer size}.
}
\label{betalayer}
\end{figure}

\searr{For completeness, we have verified that reducing the excitability within the boundary layers, while keeping $\alpha=1$, can also suppress break-up instabilities driven by discordant alternans (Fig.~\ref{betalayer}). Here, the reduction is introduced via $\tau_{r}^{0}=\beta\tau_{r}$, with $\beta$ a constant, to characterize how the enhancement of stability depends on the degree of excitability reduction. In this case, $\Delta\tau_{d}^{c}/\tau_{d}^{c}$ increases monotonically as $\beta$ decreases, until the excitability reduction is strong enough to suppress all wave activity (Fig.~\ref{betalayer}). Thus, and in contrast to the case $\alpha=0$ and $\alpha=1$, choosing $\beta=0$ or $\beta=1$ does
not result in equivalent results. Finally, larger boundary layers strengthen the stabilization, similar to the case of meandering.}

\section{Discretization and propagation failure on the $x\!-\!y$ plane}
\begin{figure}[htbp]
\includegraphics [width=8.5 cm] {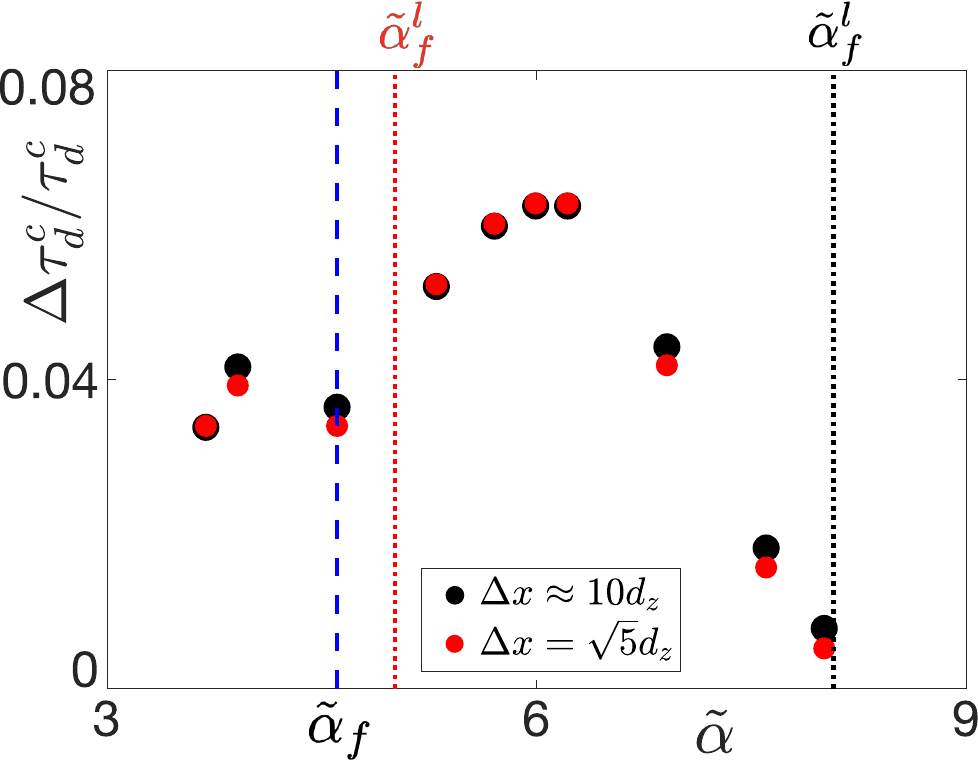} 
\caption{Effects of $\Delta x$ in the enhancement of stability of scroll waves in a slab of thickness $\mathcal{H}=8d_{z}$ with boundary layers of size $l=d_{z}$. The dashed blue line indicates  propagation failure along the transmural direction, while the dotted lines represent propagation failure in the $x-y$ plane. The $\alpha_{f}^{l}$ values were obtained from 2D simulations in a single boundary slice. As in the main text, $\tilde{\alpha}=log(\alpha)+10.82$.}
%\vspace{-18mm}
\label{FigSDx}
\end{figure}

In this section, we explore the effects of varying $\Delta x$ on the suppression of meandering instabilities. Specifically, we make the in-plane spatial grid finer, from $\approx\!\!10d_{z}$ to $\sqrt{5}d_{z}$. This particular choice results in equal conduction strength in  the $x-y$ and transmural directions. Similar to the analysis in the main text, we sweep the excitability parameter $\tau_{d}$, in steps of $0.001$ ms, starting from a rigidly rotating scroll wave and measure the corresponding enhancement of stability $(\tau_{d}^{c*}-\tau_{d}^{c})/\tau_{d}^{c}=\Delta\tau_{d}^{c}/\tau_{d}^{c}$. The results show that  the difference  
is negligible and at most
equal to the discretization step in $\tau_{d}$ (Fig.~\ref{FigSDx}).
This plot shows that the propagation failure along the transmural direction is independent of $\Delta x$ and occurs at $\alpha=\alpha_{f}$. Note, however, that the propagation failure in the $x-y$ plane within the boundary layer depends on the discretization and occurs at different $\alpha_{f}^{l}$ values (Fig.~\ref{FigSDx}). Therefore, our results suggest that the stabilization mechanism  found in the main text is  governed by transmural dynamics.

%\FloatBarrier

\section{Numerical implementation of the quasi-2D model in WebGL}
\begin{figure}[htbp]
\includegraphics [width=11.5 cm] {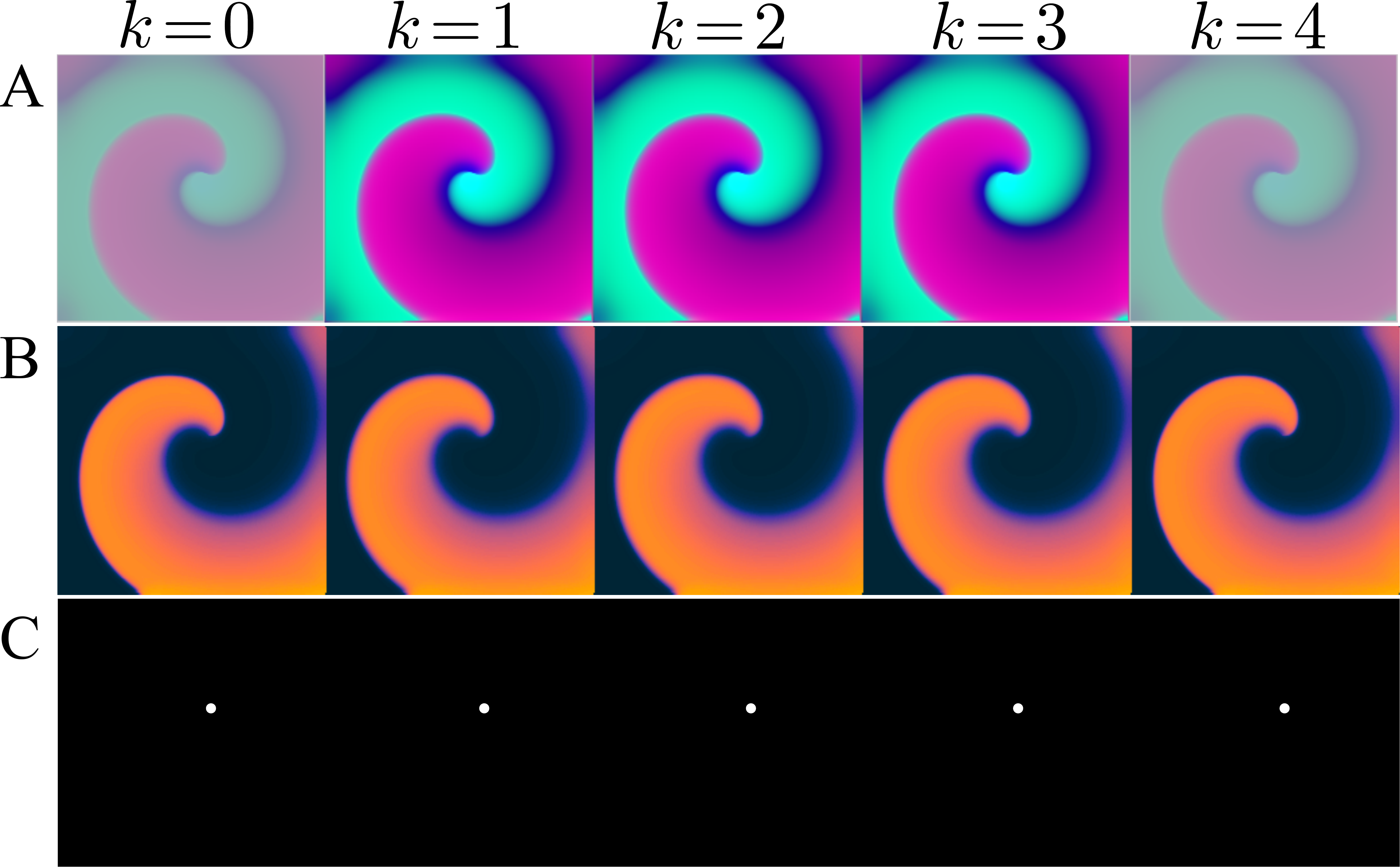} 
\caption{\sea{WebGL example for the case $\mathcal{H}=4d_z$ and $l=d_z$. (A) Initial condition in RGBA format with the $\{u,v,w\}$ fields, the $\mathcal{I}_k$ distribution, and the non-flux boundary information. (B) Canvas showing the $u$ field for $\alpha=0.02$ and $\tau_{d}=0.38$ ms. (C) Spiral tip positions of the waves in (B), enlarged for visualization purposes.}}
%\vspace{-18mm}
\label{FigWebGL}
\end{figure}

\sea{Following recent efforts to develop GPU-accelerated cardiac research codes that can be  easily used on personal laptops   \cite{kaboudian2019real,kaboudian2024fast}, we  created a WebGL implementation of the quasi-2D Fenton-Karma model studied in this work \cite{echeverria_alar_2025_codes}. Specifically, we combined the libraries Abubu.js and CCapture.js to build a GPU-accelerated code for 3D simulations in a simple geometry, capable of storing data while bypassing the usual GPU-CPU memory-transfer bottleneck. This storage protocol is possible because the WebGL simulations run on a web browser (e.g., Google Chrome, Safari, Firefox), which enables the direct implementation of an algorithm that simply takes screenshots of a canvas on the webpage. To fully exploit this capability, we solved the three-dimensional equations in an extended two-dimensional domain: the z-direction is mapped onto the x-direction (Fig.~\ref{FigWebGL}). Thus, discrete differences such as $u_{i,j,k+1}-u_{i,j,k}$ were encoded as $u_{i+L/\Delta x+1,j}-u_{i,j}$, where $i$ ($j$) indexes the discretization of the $x$($y$)-direction.}

\sea{The algorithm begins with an initial condition in RGBA format, created in Matlab, containing the transmembrane voltage $u$ and gating variables $v$ and $w$ encoded in the RGB channels, respectively (Fig.~\ref{FigWebGL}A). The transparency channel encodes the distribution of heterogeneities, $\mathcal{I}_k$, as well as the limits between z-slices in the extended domain, where non-flux boundary conditions are imposed in the $x$ and $y$ directions. Then, the WebGL code integrates Eqs.~(\ref{Eq1})-\!\!~(\ref{Eq6}) while displaying the $u$ field in real time on one canvas (Fig.~\ref{FigWebGL}B). At the same time, a numerical algorithm that computes the spiral tip position in each z-slice updates a second canvas (Fig.~\ref{FigWebGL}C). Finally, the stored data is processed using Matlab pipelines \cite{echeverria_alar_2025_codes}.}

\section{Video caption}
\textbf{Video S1}: \searr{Scroll wave stabilization through boundary layer heterogeneities in a slab of thickness $\mathcal{H}=14d_z$. The video shows the spiral tip trajectories (red circles) on the bulk slices bordering the boundary layers. Initially, the scroll wave exhibits meandering, but it stabilizes after the instantaneous decrease of $\alpha$ from 1 to $0.004$ at time $t_{het}=0.2$ s. The stabilization suppresses the meandering instability, and the scroll wave enters a rigidly rotating regime.}

\FloatBarrier
\bibliography{apssamp}% 

%\section{Video captions}
%\textbf{Video 1:} Trapping event of one scroll wave in a shell geometry generated from patient specific data of the left atrium. The video shows an extract of the temporal evolution ($SOMESECONDS$ s) of the variable $u$ (same color code as in Fig.~1 of the main text). A threshold $u_{th}  = 0.2$ (arbitrary units) is used in the video for displaying purposes only. 

%\nocite{*}